# Particle Physics and Cosmology

**Edward W. Kolb**

Fermi National Accelerator Laboratory
and The University of Chicago

In these lectures on the early Universe I will discuss some recent developments in particle cosmology, taking particular care to highlight the rôle of particle physics in our understanding of cosmology. I will assume that the reader is familiar with basic particle physics, but not necessarily basic astronomy.

Before starting, I would like to discuss the motivation for particle physicists to be interested in cosmology. The aim of modern cosmology is to understand the origin and the large-scale structure of the Universe on the basis of physical law. The modern framework for this effort is the hot big-bang model. With knowledge of the laws of physics, the fundamental forces, and the fundamental particles, in principle the model should be able to explain the gross features of our Universe. It is also possible to 'reverse engineer' this standard approach: by observations of the outcome, we might be able to tell something about the fundamental ingredients that went in. Therefore we might be able to discover something about particle physics by studying cosmology.

Let me also describe my own approach to cosmology. How should one approach the study of the history of the ancient Universe? There are two types of people who study old things, antiquarians and historians. An antiquarian is interested in things that are old simply because they are old. They do not attempt to differentiate between the relative importance of objects from antiquity. In the extreme, an antiquarian would see no difference between a grocery shopping list from 1215 and the Magna Carta. A historian on the other hand is interested in events and objects from the past because they have a bearing upon the present. It is the job of the historian to sort through the past to find the objects and events that had an impact upon the future development of history. I consider it the job of the cosmologist to be a historian of the Universe. The cosmologist should not be interested in the early Universe because it was very old, or very hot, or very dense. Rather a cosmologist studies the early Universe because he or she has the faith that events in the early Universe are responsible for shaping the present Universe, and that it is impossible to understand the Universe today without an understanding of the early Universe.

Therefore in these lectures I will concentrate on events in the early Universe that have the potential to explain the present state of the Universe. In a very real sense the job of cosmology is to provide a canvas upon which other fields of science, including particle physics, can weave their individual threads into the tapestry of our understanding of the Universe. Nowhere is the inherent unity of science better illustrated than in the interplay between cosmology, the study of the largest things in the Universe, and particle physics, the study of the smallest things.

# 1  A quick look at the Universe

Before concentrating on the particle physics aspects of cosmology, I will start with a look at the most important observational features of the Universe. I will then discuss the Robertson-Walker metric, and discuss some particle kinematics in the expanding Universe. Then I will develop the dynamics of the Friedmann-Robertson-Walker (FRW) cosmology. The final part of the introductory section will be a brief review of the radiation-dominated era and primordial nucleosynthesis. More details can be found in Kolb and Turner (1989).

## 1.1  Expansion of the Universe

It was Hubble who discovered a linear relationship between the recessional velocities of nebulae and their distances. The recessional velocity is determined via the Doppler effect. If the relative velocity between a source and observer is $v_R$, then the measured wavelength of the light, $\lambda_{\rm obs}$, will differ from the wavelength of the emitted light, $\lambda_{\rm emitted}$. This difference is expressed in terms of a redshift $z$, defined as

$$z \equiv \frac{\lambda_{\rm obs} - \lambda_{\rm emitted}}{\lambda_{\rm emitted}}. \tag{1}$$

If one interprets the observation of a redshift of light from distant galaxies as a Doppler effect, then $z = v_R/c$. (Of course this is a non-relativistic expression. The special relativistic expression relating $v_R$ and $z$ is $v_R/c = [(1+z)^2 - 1]/[(1+z)^2 + 1]$.) If the relative distance is increasing, then $z$ is positive.

The linear relationship between the distance and the redshift, Hubble's law, can be written in several equivalent forms:

$$\begin{aligned} cz &= H_0 d_L \\ v_R &= H_0 d_L \\ d_L &= (3000 h^{-1}) z \ {\rm Mpc} = 10^{-2} h^{-1} cz \ {\rm Mpc}, \end{aligned} \tag{2}$$

where a megaparsec (Mpc) is $3.1 \times 10^{24}$cm. As to the meaning of these symbols: $c$ is the speed of light (1 unless you are an astronomer), the redshift $z$ and recessional velocity $v_R$ have already been defined, $d_L$ is the (luminosity) distance, and $H_0$ is Hubble's constant. Let us postpone for a moment questions about what exactly the luminosity distance is, and just think of it as the distance to the object without worrying whether it is the distance when the light was emitted, the distance when the light was detected, *etc.*

Sixty-four years after the discovery of Hubble's law, Hubble's constant $H_0$ is still not well known. It is traditional to express the uncertainty in Hubble's law in terms of a dimensionless parameter $h$:

$$H_0 = 100\, h\, \frac{\mathrm{km}}{\mathrm{s\ Mpc}} \qquad (1 \gtrsim h \gtrsim 0.4). \tag{3}$$

The Hubble constant is *the* fundamental parameter in cosmology, and it is not known to better than a factor of two! This uncertainty traces to the oldest and most fundamental problem of astronomy—the distance scale (for a review see Rowan-Robinson, 1985). The uncertainty in $H_0$ will result in a proliferation of factors of $h$ in many of the equations in subsequent sections.

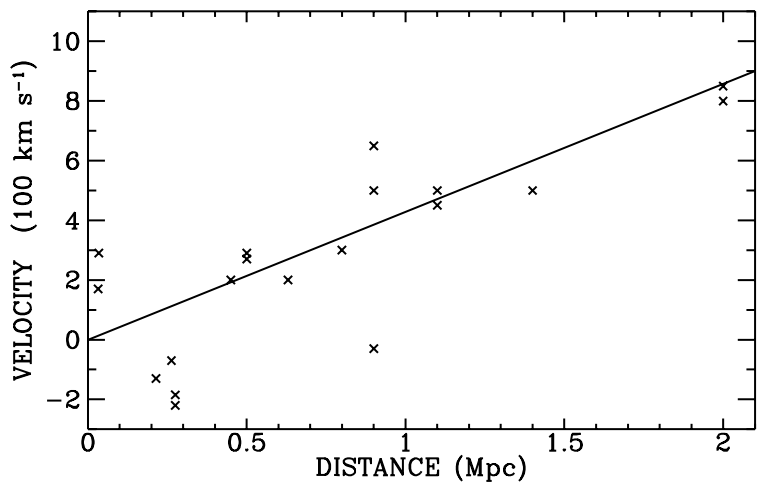

**Figure 1.** *Hubble's 1929 data. The solid line is a guide to the eye.*

It is somewhat amusing to look at the original data upon which Hubble based his claim, shown in Figure 1. Clearly it took a leap of imagination, intuition, and genius to see a linear relationship in the data. After all, some of the nearby nebulae are approaching, rather than receding. With modern (hopefully more reliable) methods for determining distances, astronomers are able to extend Hubble's program to much greater distances. All agree on the linear nature of the relationship, but do not agree on the value of $H_0$.

To orient the particle physicist I have included a small table of extragalactic distances. The distance to nearby objects in our local group of galaxies, like Andromeda, can be determined by direct means. Hence the distance is independent of $h$. For more distant objects such as the Virgo cluster of galaxies we can accurately determine the red shift (or equivalently $v_R$) but not the distance. Using the measured $z$, Hubble's law will give the distance in terms of the annoying factor of $h^{-1}$. Depending upon your favourite value of $h$, the distance to Virgo is somewhere between 12 and 30 Mpc.

| OBJECT | $z$ | $d_L$ | $v_R$ |
|---|---|---|---|
| M31 (Andromeda) | $-0.0009$ | 0.65 Mpc | $-270$ km s$^{-1}$ |
| Virgo Cluster | 0.004 | 12 $h^{-1}$Mpc | 1150 km s$^{-1}$ |
| Coma Cluster | 0.02 | 67 $h^{-1}$Mpc | 6700 km s$^{-1}$ |
| Hydra | 0.2 | 600 $h^{-1}$Mpc | 60600 km s$^{-1}$ |

Clearly Hubble's law as given in Equation 2 must break down for $z \to 1$. Even if one adopts the special relativistic Doppler formula, we will see in the section on kinematics that there are important corrections for $z \to 1$.

The expansion of the Universe and Hubble's law will be discussed further, but for our first quick view of the Universe, it will suffice to note that the Universe is expanding, and furthermore the expansion seems to be isotropic about us.

## 1.2 The cosmic background radiation

The cosmic background radiation (CBR) provides fundamental evidence that the Universe began from a hot big bang. The surface of last scattering for the CBR was the Universe at an age of about 300,000 years. The first thing to learn about the CBR is its spectrum. It is a blackbody to a remarkable accuracy. The best measurement of the spectrum of the CBR was made with the Cosmic Background Explorer (COBE) satellite (Mather *et al.* 1993). The measurements are summarized in Figure 2. Note that the true error bars for the measurements are a factor of 100 times smaller than shown in the figure. Clearly the CBR is a blackbody, with the present temperature of the Universe $T_0 = 2.726 \pm 0.01$ K, and deviations from a blackbody shape over the wavelength interval 0.05 cm to 0.5 cm less than 0.03%.

Once the temperature of the CBR is known, the number density and energy density of the background photons are also known. For a temperature of $T_0 = 2.726$K $= 2.36 \times 10^{-4}$eV, the number density and energy density of the CBR is given by

$$n_\gamma = (2\zeta(3)/\pi^2)T_0^3 = 411 \text{ cm}^{-3}$$
$$\rho_\gamma = (\pi^2/15)T_0^4 = 4.71 \times 10^{-34} \text{ g cm}^{-3}. \qquad (4)$$

After the spectrum, the next most important feature of the CBR is its isotropy. Anisotropy is expected due to several effects. For instance, a dipole moment of the CBR is expected as a result of the motion of our local reference frame with respect to the CBR rest frame. Motion with velocity $\vec{\beta} = \vec{v}/c$ through an isotropic blackbody radiation field of temperature $T$ results in a frequency-independent formula for the temperature distribution: $T(\theta) = T\sqrt{1-\beta^2}/(1-|\beta|\cos\theta)$.

The most accurate measurement of the CBR dipole anisotropy was by COBE: an amplitude of 3.336 mK, corresponding to a velocity of $627 \pm 22$ km s$^{-1}$ in the general direction of Hydra for our local group of galaxies. COBE has also determined that the dipole anisotropy has a thermal spectrum.

Additional fluctuations in the CBR temperature are also expected due to the presence of density inhomogeneities presumed to have triggered structure formation. The search for anisotropies in the CBR beyond the dipole anisotropy has occupied physicists since the discovery of the CBR itself in 1965. Finally, in 1992 the long search was rewarded when the COBE collaboration announced the discovery of anisotropy on angular scales from about 7° to 90° at a magnitude of about 1 part in $10^5$ (Smoot *et al.* 1992). There are several methods to analyze the anisotropy. The cleanest and most reliable indication of anisotropy is an *rms* temperature variation of $30 \pm 5 \mu K$ on the sky averaged over a beam of FWHM 10°. COBE also reported a quadrupole anisotropy of $11 \pm 3 \mu K$.

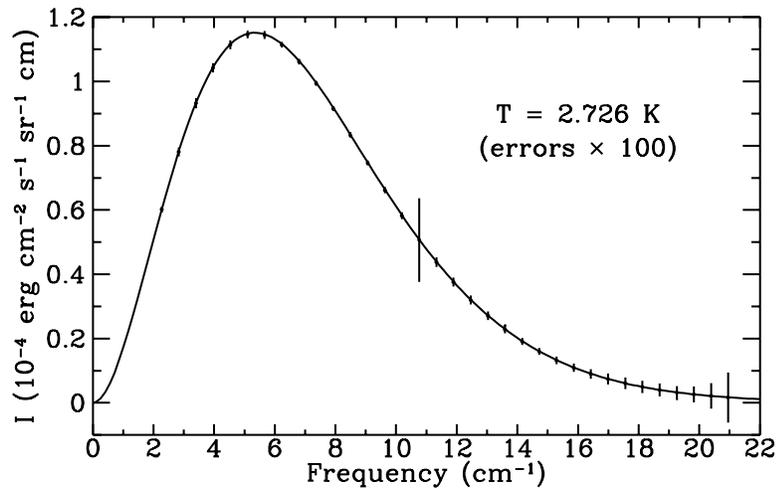

**Figure 2.** *The spectrum of the cosmic background radiation.*

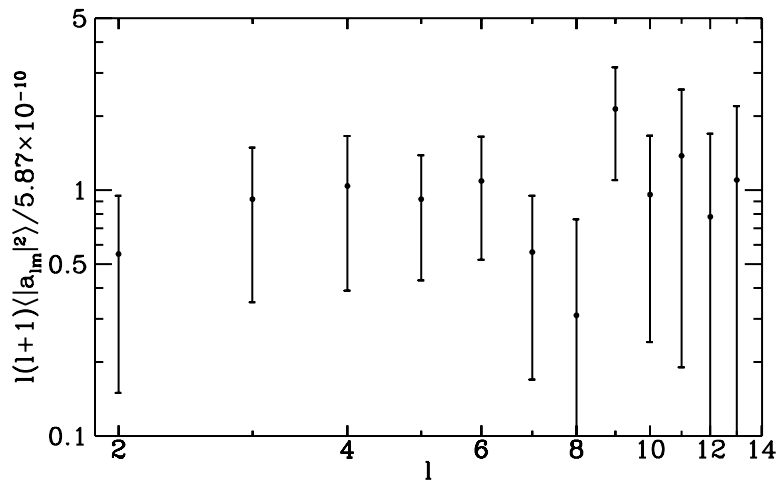

**Figure 3.** *Anisotropy multipoles of the cosmic background radiation.*

If one expands the observed temperature fluctuation as a function of angles $\theta$ and $\phi$ in spherical harmonics,

$$\frac{\delta T}{T} = \sum_{l=2}^{\infty} \sum_{m=-l}^{+l} a_{lm} Y_{lm}(\theta, \phi), \qquad (5)$$

then measurement of fluctuation can be expressed in terms of the multipole amplitudes $a_{lm}$. In this expansion the dipole moment has been left out since it arises to our peculiar velocity with respect to the CBR rest frame. The inferred multipole amplitudes from $l = 2$ to $l = 13$ as measured by COBE are shown in Figure 3.

In this first discussion of the CBR, the most important feature is that the temperature of the Universe is well determined: $T_0 = 2.726$ K. A dipole moment of the CBR is also well known. In addition there is now evidence for higher multipole moments in the CBR anisotropy. However, in the excitement and publicity of the discovery of CBR anisotropy one should not loose sight of the most important aspect of the CBR: its remarkable isotropy. The CBR is isotropic about us to better than one part in $10^5$.

## 1.3    Homogeneity and isotropy

We live in a hot, expanding Universe. We also live in a Universe that on 'large' scales is homogeneous, the same at every point, and isotropic, the same in every direction. There is an ample (and growing) body of evidence for homogeneity and isotropy. The homogeneity and isotropy of the Universe is the most fundamental principle in modern cosmology. In fact, it is called the *Cosmological Principle*.

The assumption of the isotropy and homogeneity of the Universe in modern cosmology dates back to the work of Einstein, who made the assumption not based upon observational evidence, but to simplify the mathematical analysis. Today there is ample evidence for the isotropy and homogeneity for the part of the Universe we can observe, our present Hubble volume, characterized by a length $H_0^{-1} \simeq 3000 h^{-1}$ Mpc $\simeq 10^{28} h^{-1}$ cm.

The best evidence for the isotropy of the observed Universe is the uniformity of the temperature of the CBR as discussed above. If the expansion of the Universe were not isotropic, the expansion anisotropy would lead to a temperature anisotropy of similar magnitude. Likewise, inhomogeneities in the density of the Universe on the surface of last scattering would lead to temperature anisotropies. In this regard, the CBR is a very powerful probe.

Additional evidence for the isotropy of the Universe is the isotropy of the x-ray background radiation. Some large fraction of the x-ray background is believed to be from unresolved sources (*e.g.* QSO's) at high redshift. Likewise, a substantial fraction of faint radio sources are radio galaxies at high redshift ($z \simeq 1$), and their distribution is also isotropic about us.

Evidence for homogeneity and isotropy from the distribution of galaxies is somewhat less certain, mostly because we are only now mapping the distribution of galaxies on scales large enough to see homogeneity. One can find some measure of homogeneity of the Universe by taking a sphere of radius $R$ which contains on average $N$ galaxies, and placing it down at all points in the Universe, counting the number of galaxies inside it, and computing the root-mean-square (*rms*) number fluctuations. One finds that the *rms* number fluctuations, $\delta N/N$, decrease with increasing scale, and drop below unity for radius $R_0 = 8 h^{-1}$Mpc. This indicates that on scales less than $R_0$ the Universe is lumpy, and for scales greater than $R_0$ the Universe becomes smooth.

This is not to say that the Universe is structureless on scales greater than $R_0$. The best known example of structures on larger scales comes from the Center for Astrophysics (CFA) slices of the universe (de Laupparent, Geller, and Huchra, 1986), shown in Figure 4 (recall from Equation 2 that $d_L = 10^{-2}h^{-1}cz$ Mpc). Some of the structure is the result of the presentation of the data in redshift space (which stretches out things in the radial direction), but clearly there is structure on scales much larger than $8h^{-1}$ Mpc.

Clear evidence from the distribution of galaxies for homogeneity of the Universe must await surveys on scales much larger than the CFA survey. These should be completed before the end of this century. Until that time, we can only look at larger regions of the Universe with 'sparse' samples of the location of galaxies, *i.e.* only a fraction of galaxies in the sample volume are mapped. Such a sparse sample for the Automatic Plate Measuring (APM) survey (Loveday *et al.* 1992) is shown in Figure 5. Note that the APM survey is nearly three times as deep as the CFA survey. It does not seen to show structures on the size of the survey as the CFA survey does. Just by comparing the two figures one concludes that the distribution of matter in the Universe is not a fractal, but rather approaches homogeneity on large scales.

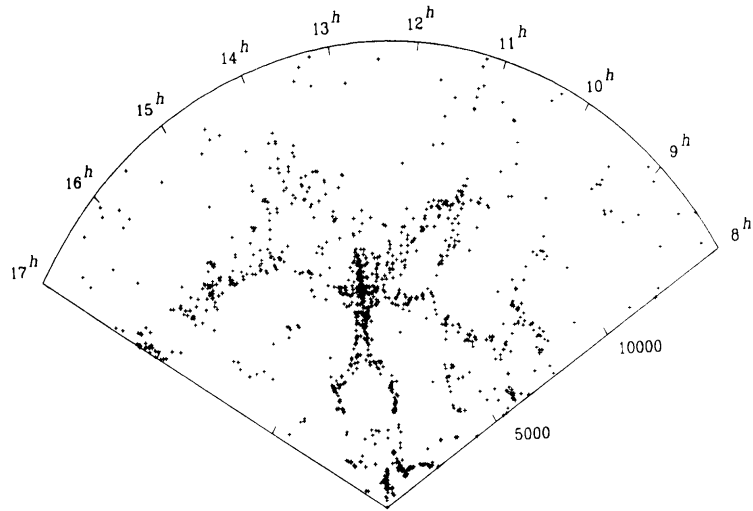

**Figure 4.** *One of the CFA slices of the Universe containing 1074 galaxies.*

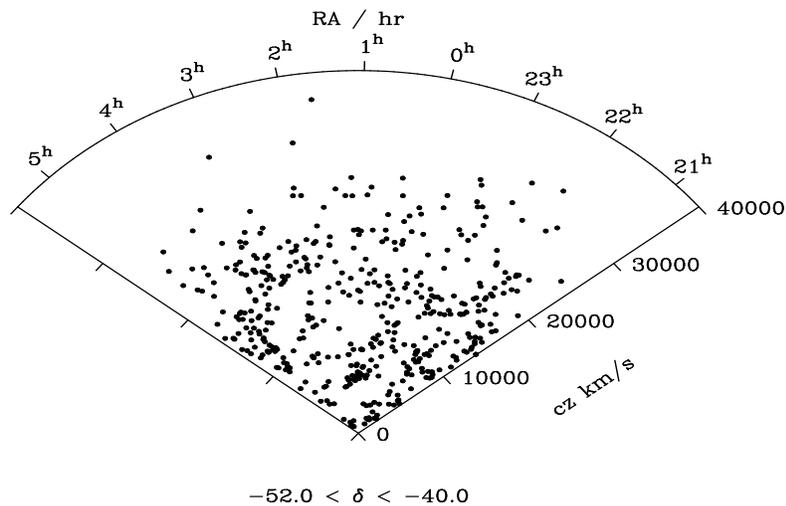

**Figure 5.** *A sparse sample of the APM survey.*

In conclusion, the Universe is lumpy on small scales, containing people, planets,

stars, galaxies, galaxy clusters, superclusters, *etc.* However on large scales the distribution of matter and radiation in the Universe is smooth. We are only now probing the transition region from lumpy to smooth in the distribution of galaxies. It is exciting to be a cosmologist at the time when the largest structures in the Universe are being discovered.

The large degree of spatial symmetry in a spatially homogeneous and isotropic Universe will greatly simplify the dynamics of the expansion of the Universe. Through the action of cosmic inflation, we will be able to understand why the Universe is homogeneous and isotropic on observable scales, as well as understanding why there is structure in the Universe.

## 1.4   The present Universe

I will conclude this quick look at the Universe today by presenting the parameters that describe the present Universe:

**Expansion:** The Universe is expanding at a present rate given by Hubble's constant, expressed in terms of a dimensionless parameter $h$: $H_0 = 100h$ km s$^{-1}$ Mpc$^{-1}$, with $0.4 < h < 1$.

**Temperature:** The present temperature of the Universe is $T_0 = 2.726 \pm 0.01$ K, with a dipole moment of 3.336 mK, and *rms* temperature fluctuations on a scale of $10°$ of about $30\mu$K.

**Homogeneity and Isotropy:** The Universe is homogeneous and isotropic on large scales and clumpy on small scales. The transition region is about $R_0 = 10h^{-1}$ Mpc.

**Mass and Energy Density:** The mass density of the Universe is poorly known. It is convenient to express mass densities in terms of a critical density, $\rho_C$, formed by Hubble's constant and Newton's constant:

$$\rho_C = \frac{3H_0^2}{8\pi G} = 1.88 h^2 \times 10^{-29} \text{g cm}^3. \tag{6}$$

Expressed as a fraction of the critical density, the matter ($M$), photon ($\gamma$), and radiation ($R$) energy densities of the Universe are

$$\begin{aligned}
\Omega_M &\equiv \rho_M/\rho_C = 0.01 \text{ to } 1 \\
\Omega_\gamma &\equiv \rho_\gamma/\rho_C = 2.6 \times 10^{-5} h^2 \\
\Omega_R &\equiv \rho_R/\rho_C = 4.3 \times 10^{-5} h^2,
\end{aligned} \tag{7}$$

where I have included 3 massless neutrino species at a temperature of 1.96 K in addition to the photons in determining the radiation energy density. Clearly today we live in a 'matter-dominated' Universe since $\Omega_M \gg \Omega_R$.

## 1.5   The Robertson-Walker metric

The metric for a space with homogeneous and isotropic spatial sections is the Robertson-Walker (RW) metric, which can be written in the form

$$ds^2 = dt^2 - a^2(t) \left\{ \frac{dr^2}{1 - kr^2} + r^2 d\theta^2 + r^2 \sin^2\theta d\phi^2 \right\} \tag{8}$$

where $(r, \theta, \phi)$ are spatial coordinates (referred to as comoving coordinates), $a(t)$ is the cosmic scale factor, and with an appropriate rescaling of the coordinates $k$ can be chosen to be $+1$, $-1$, or $0$ for spaces of constant positive, negative, or zero spatial curvature, respectively. The coordinate $r$ in Equation 8 is dimensionless, *i.e.* $a(t)$ has dimensions of length, and $r$ ranges from 0 to 1 for $k = +1$. Notice that for $k = +1$ the circumference of a one-sphere of coordinate radius $r$ in the $\phi = $ const plane is just $2\pi a(t)r$, and that the area of a two-sphere of coordinate radius $r$ is just $4\pi a^2(t)r^2$; however, the physical radius of such one- and two-spheres is $a(t)\int_0^r dr/(1-kr^2)^{1/2}$, and not $a(t)r$.

The time coordinate in Equation 8 is the proper time measured by an observer at rest in the comoving frame, *i.e.* $(r, \theta, \phi)=$const. Observers at rest in the comoving frame remain at rest, *i.e.* $(r, \theta, \phi)$ remain unchanged, and observers initially moving with respect to this frame will eventually come to rest in it. Thus, if one introduces a homogeneous, isotropic fluid initially at rest in this frame, the $t =$ const hypersurfaces will always be orthogonal to the fluid flow, and will always coincide with the hypersurfaces of both spatial homogeneity and constant fluid density.

The dynamical equations that describe the evolution of the scale factor $a(t)$ follow from the Einstein field equations, $R_{\mu\nu} - \frac{1}{2}Rg_{\mu\nu} = 8\pi G T_{\mu\nu}$. Before proceeding we must specify the stress-energy tensor. To be consistent with the symmetries of the metric, the total stress-energy tensor $T_{\mu\nu}$ must be diagonal, and by isotropy the spatial components must be equal. The simplest realization of such a stress-energy tensor is that of a perfect fluid characterized by a time-dependent energy density $\rho(t)$ and pressure $p(t)$: $T^\mu_{\ \nu} = \text{diag}(\rho, -p, -p, -p)$. The $\mu = 0$ component of the conservation of stress energy equation ($T^{\mu\nu}_{\ \ ;\nu} = 0$) gives the 1st law of thermodynamics: $d(\rho a^3) = -p d(a^3)$. For the simple equation of state $p = w\rho$, where $w$ is independent of time, $\rho$ evolves as $\rho \propto a^{-3(1+w)}$. Examples of this simple equation of state we will employ include:

$$\begin{aligned}
\text{RADIATION} \quad & (p = \frac{1}{3}\rho) \implies \rho \propto a^{-4} \\
\text{MATTER} \quad & (p = 0) \implies \rho \propto a^{-3} \\
\text{VACUUM ENERGY} \quad & (p = -\rho) \implies \rho \propto \text{const.}
\end{aligned} \quad (9)$$

The 0-0 component of the Einstein equation gives the Friedmann equation

$$\left(\frac{\dot a}{a}\right)^2 + \frac{k}{a^2} = \frac{8\pi G}{3}\rho. \qquad (10)$$

A combination of the *i-i* component with the Friedmann equation gives an equation for the deceleration of the expansion:

$$\frac{\ddot a}{a} = -\frac{4\pi G}{3}(\rho + 3p). \qquad (11)$$

The expansion rate of the Universe is determined by the Hubble *parameter* $H \equiv \dot a/a$. The Hubble parameter is not constant, and in general varies as $t^{-1}$. The Hubble time (or Hubble radius) $H^{-1}$ sets the time scale for the expansion: $a$ roughly doubles in a Hubble time. The Hubble *constant*, $H_0$, is the present value of the expansion rate. The Friedmann equation can be recast as

$$\frac{k}{H^2 a^2} = \frac{\rho}{3H^2/8\pi G} - 1 \equiv \Omega - 1. \qquad (12)$$

Since $H^2 a^2 \geq 0$, there is a correspondence between the sign of $k$, and the sign of $\Omega - 1$

$$
\begin{aligned}
k = +1 &\implies \Omega > 1 \quad \text{CLOSED} \\
k = 0 &\implies \Omega = 1 \quad \text{FLAT} \\
k = -1 &\implies \Omega < 1 \quad \text{OPEN}.
\end{aligned}
\tag{13}
$$

## 1.6  Particle kinematics

The first application of particle kinematics with the Robertson-Walker metric is a calculation of the proper distance to the horizon, *i.e.* for a comoving observer with coordinates $(r_0, \theta_0, \phi_0)$, for what values of $(r, \theta, \phi)$ would a light signal emitted at $t = 0$ reach the observer at, or before, time $t$? A light signal satisfies the geodesic equation $ds^2 = 0$. Because of the homogeneity of space, without loss of generality we may choose $r_0 = 0$. A light signal emitted from coordinate position $(r_H, \theta_0, \phi_0)$ at time $t = 0$ will reach $r_0 = 0$ in a time $t$ determined by

$$\int_0^t \frac{dt'}{a(t')} = \int_0^{r_H} \frac{dr}{\sqrt{1-kr^2}}, \tag{14}$$

and the proper distance to the horizon measured at time $t$, $d_H(t) = \int_0^{r_H} \sqrt{g_{rr}}\, dr$, is simply $a(t)$ times the above integral:

$$d_H(t) = a(t) \int_0^t \frac{dt'}{a(t')} = a(t) \int_0^{a(t)} \frac{da(t')}{\dot{a}(t')a(t')}. \tag{15}$$

We know the behaviour of $a(t)$ from the Friedmann equation. For the early Universe we can ignore the curvature term. For a radiation-dominated Universe, $a \propto t^{1/2}$, and $d_H(t) = 2t$, while for a matter-dominated Universe $a \propto t^{2/3}$, and $d_H(t) = 3t$. If $d_H(t)$ is finite, then our past light cone is limited by a particle horizon, which is the boundary between the visible Universe and the part of the Universe from which light signals have not reached us. The behaviour of $a(t)$ near the singularity will determine whether or not $d_H$ is finite.

The next application of particle kinematics is the redshift. The four-velocity $u^\mu$ of a particle with respect to the comoving frame is referred to as its *peculiar* velocity. The equation of geodesic motion for $u^\mu$ is

$$\frac{du^\mu}{d\lambda} + \Gamma^\mu_{\nu\alpha} u^\nu \frac{dx^\alpha}{d\lambda} = 0, \tag{16}$$

where $u^\mu \equiv dx^\mu/ds$, and $\lambda$ is some affine parameter, which we will choose to be the proper length $ds$. The $\mu=0$ component of the geodesic equation is $du^0/ds + \Gamma^0_{\nu\alpha} u^\nu u^\alpha = 0$. Using the fact that for the Robertson-Walker metric, the only non-vanishing component of $\Gamma^0_{\nu\alpha}$ is $\Gamma^0_{ij} = (\dot{a}/a) h_{ij}$ (where $h_{ij}$ is the spatial metric), the geodesic equation gives $|\dot{\vec{u}}|/|\vec{u}| = -\dot{a}/a$, which implies that $|\vec{u}| \propto a^{-1}$. In an expanding Universe, a freely-falling observer is destined to come to rest in the comoving frame even if he has some initial peculiar velocity. Recalling that the four-momentum is $p^\mu = mu^\mu$, we see that the magnitude of the three-momentum of a freely-propagating particle also 'redshifts' as $a^{-1}$. The wavelength of light is inversely proportional to the photon momentum ($\lambda = 2\pi\hbar/p$). If the momentum changes, the wavelength of the light also changes. The wavelength at time $t_0$, denoted as $\lambda_0$, will differ from that at time $t_1$, denoted as $\lambda_1$, by $\lambda_0/\lambda_1 = 1 + z = a(t_0)/a(t_1)$. This means that the redshift of the wavelength of a photon is due to the fact that the Universe was smaller when the photon was emitted!

Our final foray into particle kinematics will be Hubble's law. Suppose a source, *e.g.* a galaxy, has an absolute luminosity $\mathcal{L}$. Its luminosity distance is *defined* in terms of the measured flux $\mathcal{F}$ by $d_L^2 \equiv \mathcal{L}/4\pi\mathcal{F}$. If a source at comoving coordinate $r = r_1$ emits light at time $t_1$, and a detector at comoving coordinate $r = 0$ detects the light at $t = t_0$, conservation of energy implies $\mathcal{F} = \mathcal{L}/4\pi a^2(t_0)r_1^2(1+z)^2$, which implies $d_L^2 = a^2(t_0)r_1^2(1+z)^2$. In order to express $d_L$ in terms of the redshift $z$, the explicit dependence upon $r_1$ must be removed. Since light travels on geodesics,

$$\int_0^{r_1} \frac{dr}{(1-kr^2)^{1/2}} = \int_{a_1}^{a_0} \frac{da(t')}{\dot{a}(t')a(t')}. \tag{17}$$

By use of the Friedmann equation, for zero pressure solution is easily found to be

$$r_1 = \frac{2\Omega_0 z + (2\Omega_0 - 4)(\sqrt{\Omega_0 z + 1} - 1)}{H_0 R_0 \Omega_0^2 (1+z)}. \tag{18}$$

Therefore Hubble's law becomes

$$H_0 d_L = 2\Omega_0^{-2}\left[2\Omega_0 z + (2\Omega_0 - 1)\left(\sqrt{2\Omega_0 z + 1} - 1\right)\right] \simeq z + \frac{1}{2}(1-q_0)z^2 + \cdots, \tag{19}$$

where $q_0$ is the deceleration parameter, $q_0 = -\ddot{a}(t_0)/\dot{a}^2(t_0)a(t_0) = 2\Omega_0$. Clearly for $z \to 1$ departures from the linear relationship are expected. In principle these departures would lead to a value for $\Omega_0$. But in practice, evolutionary effects in the brightness of galaxies for large $z$ have prevented realization of the promise.

## 1.7 The radiation-dominated era

In a radiation-dominated Universe, the energy density and pressure can be expressed in terms of the photon temperature $T$ as

$$\rho_R = \frac{\pi^2}{30}g_* T^4, \qquad p_R = \rho_R/3 = \frac{\pi^2}{90}g_* T^4, \tag{20}$$

where $g_*$ counts the total number of effectively massless degrees of freedom (those species with mass $m_i \ll T$), given by

$$g_* = \sum_{i=\text{bosons}} g_i \left(\frac{T_i}{T}\right)^4 + \frac{7}{8}\sum_{i=\text{fermions}} g_i \left(\frac{T_i}{T}\right)^4. \tag{21}$$

The relative factor of 7/8 accounts for the difference in Fermi and Bose statistics. Note that $g_*$ is a function of $T$ since the sum runs over only those species with mass $m_i \ll T$. For $T \gtrsim 300$ GeV, all the species in the standard model—8 gluons, $W^\pm Z^o$, 3 generations of quarks and leptons, and 1 complex Higgs doublet—should have been relativistic, giving $g_* = 106.75$.

During the early radiation-dominated epoch ($t \lesssim 4 \times 10^{10}$ sec) $\rho \simeq \rho_R$; and further, when $g_* \simeq$ const, $p_R = \rho_R/3$ (*i.e.* $w = 1/3$) and $a(t) \propto t^{1/2}$. From this it follows that the expansion rate and expansion age is

$$H = 1.66 g_*^{1/2} \frac{T^2}{m_{Pl}} \ ; \qquad t = 0.301 g_*^{-1/2} \frac{m_{Pl}}{T^2} \sim \left(\frac{T}{\text{Mev}}\right)^{-2} \text{ sec.} \qquad (22)$$

## 1.8 Primordial nucleosynthesis

Primordial nucleosynthesis is a most useful probe of the early Universe and the consistency of the big bang model. The basic idea is that at very high temperatures ($T \gg 1$ MeV) there were no nuclei, but as the Universe expanded and cooled, conditions became hospitable for the formation of nuclei. The outcome of primordial nucleosynthesis, the relative abundances of the various nuclei, depend upon the interplay of the expansion rate of the Universe and the nuclear reactions. This has to occur in a setting of enormous specific entropy. In other words, primordial nucleosynthesis occurs with the ratio of photons to nucleons of about $10^9$.

The outcome of primordial nucleosynthesis is very sensitive to the baryon-to-photon ratio, usually denoted by $\eta$. The reason for this is simple. In nuclear statistical equilibrium (NSE), the fraction of the total baryon mass contributed by a nucleus with $A$ protons ($p$) and neutrons ($n$) and binding energy $B_A$ is

$$X_A = g_A [\zeta(3)^{A-1} \pi^{(1-A)/2} 2^{(3A-5)/2}] A^{5/2} (T/m_N)^{3(A-1)/2} \eta^{A-1} X_p^Z X_n^{A-Z} \exp(B_A/T), \qquad (23)$$

where as usual,

$$\eta \equiv \frac{n_N}{n_\gamma} = 2.68 \times 10^{-8} \, (\Omega_B h^2) \qquad (24)$$

is the present baryon-to-photon ratio, $g_A$ is the number of spin degrees of freedom of the nucleus, and $m_N$ is the mass of a nucleon. The sensitive dependence upon of primordial nucleosynthesis upon $\eta$ arises from the factor of $\eta^{A-1}$ in this equation. Of course primordial nucleosynthesis represents a departure from nuclear statistical equilibrium and the NSE abundance is not always the actual abundance, but nevertheless the NSE abundance sets the value of what the abundance 'wants to be'.

Recent analysis (Walker *et al.* 1991) of the outcome of primordial nucleosynthesis suggest that $\eta$ must be in the range of 3 to 5 times $10^{-10}$ to agree with the inferred primordial abundances.

## 2 The formation of structure

Before discussing the physical processes important in the theory of structure formation through gravitational instability, I will briefly review some preliminaries related to a Fourier analysis of the density field of the Universe.

It is convenient to discuss the density field of the Universe in terms of the density contrast, where

$$\delta(\vec{x}) \equiv \frac{\delta \rho(\vec{x})}{\overline{\rho}} = \frac{\rho(\vec{x}) - \overline{\rho}}{\overline{\rho}}, \qquad (25)$$

and to express the density contrast $\delta(\vec{x})$ in a Fourier expansion:

$$\delta(\vec{x}) = \frac{V}{(2\pi)^3} \int \delta_k \exp(-i\vec{k}\cdot\vec{x}) d^3k . \tag{26}$$

Here $\overline{\rho}$ is the average density of the Universe, periodic boundary conditions have been imposed, and $V$ is a normalization volume. Since $\delta(\vec{x})$ is a scalar quantity, one can use either comoving or physical coordinates in the Fourier expansion; unless otherwise specified, I will use comoving coordinates.

A particular Fourier component is characterized by its amplitude $|\delta_k|$ and its comoving wavenumber $k$. Since $\vec{x}$ and $\vec{k}$ are (comoving) coordinate quantities, the physical distance and physical wavenumber are related to the comoving distance and wavenumber by $dx_{\rm phys} = a(t)dx$, $k_{\rm phys} = k/a(t)$. The wavelength of a perturbation is related to the wavenumber by $\lambda \equiv 2\pi/k$, $\lambda_{\rm phys} = a(t)\lambda$.

All statistical quantities for *gaussian* random fluctuations can be specified in terms of the *power spectrum* $|\delta_k|^2$. In the absence of a better idea it is assumed that $|\delta_k|^2 \propto k^n$, that is, a featureless power law.

The *rms* density fluctuation is defined by,

$$\frac{\delta\rho}{\rho} = \langle \delta(\vec{x})\delta(\vec{x})\rangle^{1/2}, \tag{27}$$

where $\langle\cdots\rangle$ indicates the average over all space. Some manipulation yields:

$$\left(\frac{\delta\rho}{\rho}\right)^2 = V^{-1}\int_0^\infty \frac{k^3|\delta_k|^2}{2\pi^2}\frac{dk}{k}. \tag{28}$$

The contribution to $(\delta\rho/\rho)^2$ from a given logarithmic interval in $k$ is given by

$$\left(\frac{\delta\rho}{\rho}\right)^2_k \approx \Delta^2(k) \equiv V^{-1}\frac{k^3|\delta_k|^2}{2\pi^2}. \tag{29}$$

The fluctuation power per logarithmic interval, denoted by $\Delta^2(k)$, will appear often.

Now consider $(\delta M/M)$, the *rms* mass fluctuation on a given mass scale. This is what most people mean when they refer to the density contrast on a given mass scale. Mechanically, one would measure $(\delta M)_{rms}$ as follows: Take a volume $V_W$, which on average contains mass $M$, place it at all points throughout space, measure the mass within it, and then compute the *rms* mass fluctuation. Although it is simplest to choose a spherical volume $V_W$ with a sharp surface, to avoid surface effects one often wishes to smear the surface. This is done by using a *window function* $W(r)$, which smoothly defines a volume $V_W$ and mass $M = \overline{\rho}V_W$, where

$$V_W = 4\pi \int_0^\infty r^2 W(r) dr. \tag{30}$$

The *rms mass* fluctuation on the mass scale $M \equiv \overline{\rho}V_W$ is given in terms of the density contrast and window function by

$$\left(\frac{\delta M}{M}\right)^2 = \frac{1}{V_W^2}\int \Delta^2(k)|W(k)|^2\frac{dk}{k}. \tag{31}$$

Notice that the *rms* mass fluctuation is given in terms of an integral over $\Delta^2(k)$.

Taking $|\delta_k|^2 = A k^n$, and $W(r) = 1$ for $r \leq r_0$ and zero otherwise, one finds

$$\left(\frac{\delta M}{M}\right)_{r_0} \simeq \Delta(k = r_0^{-1}) \tag{32}$$

for $n > -3$. $(\delta M/M)$ is given by an integral over all wavelengths longer than about $r_0$ and is roughly equal to the *rms* value of $\Delta(k = r_0^{-1})$. Using this 'top hat' window function, Davis and Peebles (1983) find for the CFA-I redshift survey that $(\delta M/M) = 1$ for a sphere of radius of $r_0 = 8h^{-1}$ Mpc. This value of $r_0$ separates the linear from the non-linear regime.

It is commonly assumed that the observed structure is a result of the growth of small seed inhomogeneities. In the next section on inflation and in Section 4 on phase transitions I will discuss possible origins for the seed perturbations. But before doing that, here I will discuss the theory of gravitational instability in an expanding Universe, and discuss the physical, non-gravitational, processes that might affect the perturbation spectrum. First, consider the linear theory of gravitational instability.

## 2.1    Gravitational instability—linear theory

We will start by considering the simplest possible form of gravitational instability, the Jeans instability in a non-expanding, perfect fluid. The Eulerian equations of Newtonian motion describing a perfect fluid are

$$\begin{aligned}
&\frac{\partial \rho}{\partial t} + \vec{\nabla} \cdot (\rho \vec{v}) = 0, \\
&\frac{\partial \vec{v}}{\partial t} + (\vec{v} \cdot \vec{\nabla})\vec{v} + \frac{1}{\rho}\vec{\nabla}p + \vec{\nabla}\phi = 0, \\
&\nabla^2 \phi = 4\pi G \rho.
\end{aligned} \tag{33}$$

Here $\rho$ is the matter density, $p$ the matter pressure, $\vec{v}$ the local fluid velocity, and $\phi$ the gravitational potential. The simplest solution is the static one where the matter is at rest ($\vec{v}_0 = 0$) and uniformly distributed in space ($\rho_0 = $ const, $p_0 = $ const). Throughout, we will denote unperturbed quantities with the subscript 0. Now consider perturbations about this static solution, expanded as

$$\rho = \rho_0 + \rho_1; \quad p = p_0 + p_1; \quad \vec{v} = \vec{v}_0 + \vec{v}_1; \quad \phi = \phi_0 + \phi_1. \tag{34}$$

We will consider adiabatic perturbations, that is, perturbations for which there are no spatial variations in the equation of state. The (adiabatic) sound speed, $v_S^2$, is defined as $v_S^2 \equiv \partial p / \partial \rho$, and by assumption there are no spatial variations in the equation of state, $v_S^2 = p_1/\rho_1$. To first order, the small perturbation $\rho_1$ satisfies the second-order differential equation:

$$\frac{\partial^2 \rho_1}{\partial t^2} - v_S^2 \nabla^2 \rho_1 = 4\pi G \rho_0 \rho_1. \tag{35}$$

Solutions to Equation 35 are of the form

$$\rho_1(\vec{r}, t) = \delta(\vec{r}, t)\rho_0 = A \exp\left[-i\vec{k}\cdot\vec{r} + i\omega t\right]\rho_0, \tag{36}$$

and $\omega$ and $\vec{\mathbf{k}}$ satisfy the dispersion relation $\omega^2 = v_S^2 k^2 - 4\pi G \rho_0$, with $k \equiv |\vec{\mathbf{k}}|$.

If $\omega$ is imaginary, there will be exponentially growing modes; if $\omega$ is real, the perturbations will simply oscillate as sound waves. It is clear that for $k$ less than some critical value, $\omega$ will be imaginary. This critical value is called the Jeans wavenumber, $k_J$, and is given by

$$k_J = \left(\frac{4\pi G \rho_0}{v_S^2}\right)^{1/2}. \qquad (37)$$

For $k^2 \ll k_J^2$, $\rho_1$ grows exponentially on the dynamical timescale $\tau_{\rm dyn} \simeq (4\pi G \rho_0)^{-1/2}$.

It is useful to define the Jeans mass, the total mass contained within a sphere of radius $\lambda_J/2 = \pi/k_J$:

$$M_J \equiv \frac{4\pi}{3}\left(\frac{\pi}{k_J}\right)^3 \rho_0 = \frac{\pi^{5/2}}{6}\frac{v_S^3}{G^{3/2}\rho_0^{1/2}}. \qquad (38)$$

Perturbations of mass less than $M_J$ are stable against gravitational collapse, while those of mass greater than $M_J$ are unstable.

The classical Jeans analysis is not directly applicable to cosmology because the expansion of the Universe is not taken into account, and because the analysis is Newtonian. For modes of wavelength less than that of the horizon, i.e. $\lambda_{\rm phys} \ll H^{-1}$, a Newtonian analysis suffices so long as the expansion is taken into account.

When the expansion of the Universe is taken into account, the wave equation becomes

$$\ddot{\delta}_k + 2\frac{\dot{a}}{a}\dot{\delta}_k + \left(\frac{v_S^2 k^2}{a^2} - 4\pi G \rho_0\right)\delta_k = 0. \qquad (39)$$

Again, for $k \ll k_J$ there are unstable (growing mode) solutions. In the limit $k \ll k_J$ for a spatially flat, matter-dominated FRW model where $\dot{a}/a = (2/3)t^{-1}$ and $\rho_0 = (6\pi G t^2)^{-1}$,

$$\ddot{\delta} + \frac{4}{3t}\dot{\delta} - \frac{2}{3t^2}\delta = 0, \qquad k \ll k_J. \qquad (40)$$

This equation has two independent solutions, a growing mode, $\delta_+$, and a decaying mode, $\delta_-$, with time dependence given by

$$\delta_+(t) = \delta_+(t_i)\left(\frac{t}{t_i}\right)^{2/3}; \quad \delta_-(t) = \delta_-(t_i)\left(\frac{t}{t_i}\right)^{-1}. \qquad (41)$$

Here we see the key difference between the Jeans instability in the static regime and in the expanding Universe: the expansion of the Universe slows the exponential growth of the instability and results in power-law growth for unstable modes.

Consider a two-component model with non-relativistic (NR) species $i$ (e.g. baryons or a WIMP—weakly interacting massive particle) and photons during the radiation-dominated era. In this case $\dot{a}/a = 1/2t$ and $\rho_0 \ll \rho_{\rm TOTAL}$. Consider the evolution of perturbations that are Jeans unstable, $k \ll k_J$. If the photons have no perturbations then the solution is

$$\ddot{\delta}_i + \frac{1}{t}\dot{\delta}_i = 0. \qquad (42)$$

In this case the solution is $\delta_i(t) = \delta_i(t_i)[1 + a \ln(t/t_i)]$, so only a perturbation with an 'initial velocity,' $\dot{\delta}(t_i) \neq 0$, can actually grow, and only logarithmically at that.

The growth of linear perturbations in a radiation-dominated Universe is inhibited compared to the static situation. The physical reason behind this fact is easy to see. The classical Jeans instability with exponential growth is moderated by the expansion of the Universe. In a matter-dominated epoch, perturbations grow as a power law. In the radiation-dominated epoch, the expansion rate is faster than what it would be if there were only matter present, and so the growth of perturbations is further slowed.

## 2.2 Damping processes

The theory of gravitational instability discussed so far assumes that the Universe is filled with a perfect fluid. However there are important departures from this ideal situation.

Collisionless damping occurs during the radiation-dominated era when (linear) perturbations do not grow. If the particle species is collisionless the perfect-fluid approximation is clearly invalid. In this case collisionless phase mixing, or Landau damping, will occur. Perturbations will be damped on length scales smaller than the distance the particle will travel while decoupled. If the species becomes non-relativistic at time $t_{\rm NR}$, then at time $t_{\rm EQ}$ when the Universe becomes matter-dominated and perturbations can start to grow the species would have free-streamed a distance

$$\lambda_{FS} = a(t) \int_0^{t_{\rm NR}} \frac{1}{a(t')} dt' + a(t) \int_{t_{\rm NR}}^{t_{\rm EQ}} \frac{v(t')}{a(t')} dt', \qquad (43)$$

where the integral has been split into two pieces: the relativistic regime, with $v \simeq 1$; and the non-relativistic regime, when $v \lesssim 1$. Assuming the Universe is radiation dominated at $t_{\rm NR}$,

$$\lambda_{FS} \simeq t_{\rm NR} (1 + z_{\rm NR}) [2 + \ln(t_{\rm EQ}/t_{\rm NR})]. \qquad (44)$$

For a light neutrino species

$$\lambda_{FS-\nu} \simeq 20 \text{ Mpc} \left(\frac{m_\nu}{30 \text{eV}}\right)^{-1}. \qquad (45)$$

Perturbations on scales less than $\lambda_{FS-\nu}$ are damped by free streaming. Note that this length scales is much larger than the length scale associated with galaxies, containing a mass (in neutrinos) of $4 \times 10^{14} (m_\nu/30\text{eV})^{-2} M_\odot$, where $M_\odot = 2 \times 10^{33}$g is a solar mass.

The perfect-fluid approximation also breaks down for baryons. Before recombination the photons mean free path is small, but as the matter particles in Universe become electrically neutral during recombination the photon mean free path becomes longer, and photons can diffuse out of dense regions. To the degree that the photons are not completely decoupled from the baryons, they can drag the baryons along, also damping perturbations in the baryons. This effect is known as Silk damping. A careful calculation of the damping requires solving the Boltzmann equation, but to a good approximation the scale for Silk damping is about an order of magnitude smaller than the horizon scale at decoupling. Thus, baryon perturbations should be damped on scales less than $\lambda_S \simeq (\Omega_0/\Omega_B)^{1/2}(\Omega_0 h^2)^{-3/4}$ Mpc, corresponding to a mass of $M_S = 6.2 \times 10^{12}(\Omega_0/\Omega_B)^{3/2}(\Omega_0 h^2)^{-5/4} M_\odot$.

## 2.3 Super-horizon-size perturbations

So far the analysis of the evolution of density perturbations has been Newtonian. For modes that are well within the horizon, $\lambda_{\text{phys}} \ll H^{-1}$, the Newtonian analysis is adequate. To treat the evolution of modes outside the horizon one needs a full, general-relativistic analysis. Clearly this is beyond the scope of these lectures. However I will illustrate the idea in a simple, geometric way. Consider perturbations of a spatially flat ($k = 0$) FRW model. The Friedmann equation for the unperturbed $k = 0$ model is

$$H^2 = 8\pi G \rho_0 / 3 \qquad (k = 0). \tag{46}$$

Now consider a similar FRW model, one with the same expansion rate $H$, but one that has higher density, $\rho = \rho_1$, and is therefore positively curved. The expansion rate is

$$H^2 = \frac{8\pi G \rho_1}{3} - \frac{k}{a^2} \quad (k > 0). \tag{47}$$

If we compare the models when their expansion rates are equal, we have made a choice of gauge; in this case the uniform Hubble gauge. We immediately see that the density contrast between the two models is given in terms of the curvature of the closed model by

$$\delta \equiv \frac{\rho_1 - \rho_0}{\rho_0} = \frac{k/a^2}{8\pi G \rho_0 / 3}. \tag{48}$$

The evolution of $\delta$ has been reduced to that of the curvature $k/a^2$ relative to the energy density $\rho_0$. As long as $\delta$ is small, equivalently $k/a^2 \lesssim 8\pi G \rho_0$, the scale factors for the two models are essentially equal (fractional difference of order $\delta$). In a matter-dominated Universe, $\rho \propto a^{-3}$, while in a radiation-dominated Universe $\rho \propto a^{-4}$, so

$$\delta \propto \frac{a^{-2}}{\rho_0} \propto \begin{cases} a^2 & \text{RADIATION DOMINATED} \\ a & \text{MATTER DOMINATED}. \end{cases} \tag{49}$$

Recall that in a matter-dominated Universe $a \propto t^{2/3}$, while in a radiation-dominated Universe $a \propto t^{1/2}$, so

$$\delta = \delta_i \begin{cases} t/t_i & \text{RADIATION DOMINATED} \\ (t/t_i)^{2/3} & \text{MATTER DOMINATED}. \end{cases} \tag{50}$$

This simple model illustrates several important points about the evolution of super-horizon-sized perturbations. (1) The geometric character of a density perturbation, which is why density perturbations are referred to as curvature fluctuations. (2) What is actually relevant is the difference in the evolution of the perturbed model as compared to some unperturbed, reference model. (3) The freedom in the choice of the reference model is equivalent to a gauge choice, so that in general, $\delta$ will be gauge dependent.

As in particle physics, when confronted with gauge ambiguity, the correct thing to do is to ask a physical question, one whose answer cannot depend upon the gauge. Here, the perturbed and reference models are compared by matching their expansion rates. A very useful quantity is $\zeta \equiv \delta\rho/(\rho_0 + p_0)$. The evolution of $\zeta$ is particularly simple and *independent* of the background space-time. For super-horizon-sized modes the evolution is $\zeta = \text{const}$ for $\lambda_{\text{phys}} \gtrsim H^{-1}$.

## 2.4   Summary of the evolution of perturbations

For sub-horizon-sized perturbations one can perform a Newtonian treatment of the evolution of perturbations. During the radiation-dominated epoch, perturbations do not grow. During the matter-dominated epoch, perturbations on scales larger than the Jeans length grow as $\delta \propto a(t) \propto t^{2/3}$. Perturbations on scales smaller than the Jeans length oscillate as acoustic waves. Before recombination, the baryon Jeans mass is larger than the horizon mass, by a factor of about 30; after recombination the baryon Jeans mass drops to about $10^5 M_\odot$.

Collisionless phase mixing damps perturbations on scales smaller than the free-streaming scale, $\lambda_{FS}$ is a few times $(t_{NR}/a_{NR})$, where the subscript NR denotes the value of the quantity at the epoch when the species became non relativistic. Taking the particle's mass to be $m_X$ and the ratio of its temperature to the photon temperature to be $T_X/T$, the free-streaming scale is roughly $\lambda_{FS} \simeq 1\text{ Mpc }(m_X/\text{keV})^{-1}(T_X/T)$. For cold dark matter (CDM) models this damping scale is smaller than any cosmologically interesting scale. For hot dark matter (HDM) models the damping scale is larger than the galactic scale, so HDM must be augmented with some other seeds to grow galaxies.

Due to photon diffusion, adiabatic fluctuations in the baryons are strongly damped on scales smaller than the Silk scale. This damping occurs primarily just as the photons and baryons decouple. The Silk scale is given by $\lambda_S \simeq 3.5(\Omega_0/\Omega_B)^{1/2}(\Omega_0 h^2)^{-3/4}$ Mpc.

For modes that are super-horizon sized the subtleties of the gauge non-invariance of $\delta\rho$ are important, and a full general relativistic treatment is required. There are two physical modes, a decaying mode and a growing mode, as well as pure gauge modes. In the synchronous gauge, the growing mode evolves as $\delta \propto a^2 \propto t$ (radiation dominated) and $\delta \propto a(t) \propto t^{2/3}$ (matter dominated). Alternatively, the evolution of super-horizon-sized modes can be described by the quantity $\zeta = \delta\rho/(\rho_0 + p_0)$, which is constant.

The primeval spectrum is modified by the above physical processes. The processing of the initial spectrum by the damping processes depend upon the mix of matter: cold, hot, and baryons. It is useful to quantify the processing of the spectrum by specifying a 'transfer function'. Since matter fluctuations start to grow when the Universe becomes matter dominated, it is convenient to specify the spectrum at this time, $t_{\text{EQ}} = 4 \times 10^{10}(\Omega_0 h^2)^{-2}$sec. Again, in the absence of a better idea, it is convenient to specify the *unprocessed* power spectrum as a power law, $|\delta_k|^2 \propto k^n$.

The processed spectrum for hot dark matter is given by

$$|\delta_k|^2 \;=\; Ak^n 10^{-2(k/k_\nu)^{1.5}} \;=\; Ak^n \exp[-4.61(k/k_\nu)^{1.5}], \tag{51}$$

where the neutrino damping scale is $k_\nu = 0.16(m_\nu/30\text{eV})\text{Mpc}^{-1}$ (which is equivalent to $\lambda_\nu = 40(m_\nu/30\text{eV})^{-1}\text{Mpc}$), and $A$ provides the overall normalization of the power spectrum. For cold dark matter the processed spectrum is given by

$$|\delta_k|^2 = \frac{Ak^n}{(1 + \beta k + \omega k^{1.5} + \gamma k^2)^2}, \tag{52}$$

with $\beta = 1.7(\Omega_0 h^2)^{-1}$ Mpc, $\omega = 9.0(\Omega_0 h^2)^{-1.5}$ Mpc$^{1.5}$, $\gamma = 1.0(\Omega_0 h^2)^{-2}$ Mpc$^2$. Of course there is an intermediate case known as warm dark matter.

| OBJECT | MASS | LENGTH SCALE | ANGULAR SCALE |
|---|---:|---:|---:|
| Stars | $1 M_\odot$ | $0.0004 h^{-2/3}$ Mpc | $0.0065'' h^{1/3}$ |
| Globular Clusters | $10^6 M_\odot$ | $0.04 h^{-2/3}$ Mpc | $0.65'' h^{1/3}$ |
| Galaxies | $10^{12} M_\odot$ | $2 h^{-2/3}$ Mpc | $65'' h^{1/3}$ |
| Groups of Galaxies | $10^{13} M_\odot$ | $4 h^{-2/3}$ Mpc | $2.2' h^{1/3}$ |
| Non-Linear Scale | $10^{14} h^{-1} M_\odot$ | $8 h^{-1}$ Mpc | $4.3'$ |
| Thickness of LSS | $5 \times 10^{14} h^{-1} M_\odot$ | $15 h^{-1}$ Mpc | $8.5'$ |
| Clusters | $10^{15} M_\odot$ | $20 h^{-2/3}$ Mpc | $11' h^{1/3}$ |
| Superclusters | $10^{16} M_\odot$ | $40 h^{-2/3}$ Mpc | $23' h^{1/3}$ |
| Horizon at LSS | $10^{18} h^{-1} M_\odot$ | $200 h^{-1}$ Mpc | $2°$ |

**Table 1.** *Length and angular scales in an $\Omega_0 = 1$ Universe. The mass is related to scale by $M(\lambda) = 1.45 \times 10^{11} h^2 \lambda_{\mathrm{Mpc}}^3 M_\odot$, and the angle is related to scale by $\Delta\theta = 34'' h \lambda_{\mathrm{Mpc}}$. LSS stands for last scattering surface, $z = 1100$.*

## 2.5 Confronting the spectrum

In the next two sections we will discuss the generation of perturbations in inflation and due to defects produced in phase transitions. He we discuss how we can probe the spectrum by present-day observations. For more details, see Liddle and Lyth (1993) or Copeland, Kolb, Liddle, and Lidsey (1993b).

The range of scales of interest stretches from the present horizon scale, $6000 h^{-1}$ Mpc, down to about $1 h^{-1}$ Mpc, the scale which contains roughly enough matter to form a typical galaxy. On the microwave sky, an angle of $\theta$ (for small enough $\theta$) samples linear scales of $100 h^{-1}(\theta/1°)$Mpc. For purposes of discussion, it is convenient to split this range into three separate regions.

- **A: Large scales: $6000 h^{-1}$ Mpc $\longrightarrow$ $\sim 200 h^{-1}$ Mpc:**
  These scales entered the horizon after the decoupling of the microwave background. Except in models with peculiar matter contents, perturbations on these scales have not been affected by any physical processes, and the spectrum retains its original form. At present the perturbations are still very small, growing in the linear regime without mode coupling. Here, we are still seeing the primeval spectrum.

- **B: Intermediate scales: $\sim 200 h^{-1}$ Mpc $\longrightarrow$ $8 h^{-1}$ Mpc:**
  These scales remain in the linear regime, and their gravitational growth is easily calculable. However, they have been seriously influenced by the matter content of the Universe, in a way normally specified by a *transfer function*, which measures the decrease in the density contrast relative to the value it would have had if the primeval spectrum had been unaffected. Even in CDM models, where the only effect is the suppression of growth due to the Universe not being completely matter dominated at the time of horizon entry, this effect is at the 25% level at 200$h^{-1}$ Mpc. To reconstruct the primeval spectrum on these scales, it is thus essential to know the matter content of the Universe, including dark matter, and of its influence on the growth of density perturbations.

- **C: Small scales:** $8h^{-1}$ **Mpc** $\longrightarrow$ $1h^{-1}$ **Mpc:**
  On these scales the density contrast has reached the nonlinear regime, coupling together modes at different wavenumbers, and it is no longer easy to calculate the evolution of the density contrast. This can be attempted either by an approximation scheme such as the Zel'dovich approximation (Efstathiou, 1990), or more practically via $N$-body simulations (for example, see Davis, *et al.* 1992, and references therein). Further, hydrodynamic effects associated with the nonlinear behaviour can come into play, giving rise to an extremely complex problem with important non-gravitational effects. Again, the transfer function plays a crucial role on these scales. In hot dark matter models, perturbations on these scales are most likely almost completely erased by free streaming, and hence no information can be expected to be available.

Let us now consider each range of scales in turn, starting with the largest scales and working down to the smallest scales.

## A. Large scales ($6000h^{-1}$ Mpc $\longrightarrow$ $\sim 200h^{-1}$ Mpc):

Without doubt the most important form of observation on large scales for the near future is large-angle microwave background anisotropies. Scales of a couple of degrees or more fall into our definition of large scales. Such measurements are of the purest form available—anisotropy experiments directly measure the gravitational potential at different parts of the sky, on scales where the spectrum retains its primeval form. Such measurements also are of interest in that the tensor modes may contribute.

In addition to perturbations from the scalar density perturbations, the presence of gravitational waves will lead to temperature fluctuations. One can think of gravitational waves as propagating modes associated with transverse, traceless tensor metric perturbations of $g_{\mu\nu} \to g_{\mu\nu}^{\rm FRW} + h_{\mu\nu}$.

Tensor modes do not participate in structure formation and most measurements we shall discuss are oblivious to them. Further, tensor modes inside the horizon redshift away relative to matter, and so tensor modes also fail to participate in small-angle microwave background anisotropies.

Nevertheless, these large-scale measurements still exhibit one crucial and ultimately uncircumventable problem. On the largest scales, the number of statistically independent sample measurements that can be made is small. Given that the underlying inflationary fluctuations are stochastic, one obtains only a limited set of realizations from the complete probability distribution function. Such a subset may insufficiently specify the underlying distribution. This effect, which has come to be known as the *cosmic variance*, is an important matter of principle, being a source of uncertainty which remains even if perfectly accurate experiments could be carried out. At any stage in the history of the Universe, it is impossible to accurately specify the properties (most significantly the mean, which is what the spectrum specifies assuming gaussian statistics) of the probability distribution function pertaining to perturbations on scales close to that of the observable Universe.

Observations other than microwave background anisotropies appear confined to the long term future. Even such an ambitious project as the Sloan Digital Sky Survey (SDSS) (Gunn and Knap 1992; Kron 1992) can only reach out to perhaps $500h^{-1}$ Mpc, which can only touch the lower end of our specified large scales. However, in order to specify the fluctuations accurately, one needs many statistically independent regions (100 seems an optimistic lower estimate) which means that the SDSS may not specify the spectrum with sufficient accuracy above perhaps $100h^{-1}$ Mpc.

A much more crucial issue is that the SDSS will measure the galaxy distribution power spectrum, not the mass distribution power spectrum. In modern work it is taken almost completely for granted that these are not the same, and it seems likely too that a bias parameter (relating the two by a multiplicative constant) which remains scale independent over a wide range of scales may be hopelessly unrealistic. Consequently, converting from the galaxy power spectrum back to that of the matter may require a detailed knowledge of the process of galaxy formation and the environmental factors around distant galaxies. Once one attempts to reach yet further galaxies with a long look-back time, one must also understand something about evolutionary effects on galaxies. As we shall discuss in the section on intermediate scales, it seems likely that peculiar velocity data may be rather more informative than the statistics of the galaxy distribution.

A more useful tool for large scales is microwave background anisotropies on large angular scales. Our formalism closely follows that of Scaramella and Vittorio (1990). On large angular scales, the most convenient tool for studying microwave background anisotropies is the expansion into spherical harmonics

$$\frac{\Delta T}{T}(\vec{x}, \theta, \phi) = \sum_{l=2}^{\infty} \sum_{m=-l}^{l} a_{lm}(\vec{x})\, Y_m^l(\theta, \phi), \qquad (53)$$

where $\theta$ and $\phi$ are the usual spherical polar angles and $\vec{x}$ is the observer position. With spherical harmonics defined as in Press *et al.* 1986, the reality condition is $a_{l,-m} = (-1)^m\, a_{l,m}^*$. In the expansion, the unobservable monopole term has been removed. The dipole term has also been completely subtracted; the intrinsic dipole on the sky cannot be separated from that induced by our peculiar velocity relative to the comoving frame.

With gaussian statistics for the density perturbations, the coefficients $a_{lm}(\vec{x})$ are gaussian distributed stochastic random variables of position, with zero mean and rotationally invariant variance depending only on $l$: $\langle a_{lm}(\vec{x}) \rangle = 0$; $\langle |a_{lm}(\vec{x})|^2 \rangle \equiv \Sigma_l^2$.

It is crucial to note that a single observer sees a single realization from the probability distribution for the $a_{lm}$. The observed multipoles as measured from a single point are defined as

$$Q_l^2 = \frac{1}{4\pi} \sum_{m=-l}^{l} |a_{lm}|^2, \qquad (54)$$

and indeed the temperature autocorrelation function can be written in terms of these

$$C(\alpha) \equiv \left\langle \frac{\Delta T}{T}(\theta_1, \phi_1)\frac{\Delta T}{T}(\theta_2, \phi_2) \right\rangle_\alpha = \sum_{l=2}^\infty Q_l^2 P_l(\cos \alpha), \qquad (55)$$

where the average is over all directions on a single observer sky separated by an angle $\alpha$, and $P_l(\cos \alpha)$ is a Legendre polynomial. The expectation for the $Q_l^2$, averaged over all observer positions, is just $4\pi \langle Q_l^2 \rangle = (2l+1)\Sigma_l^2$.

A given model predicts values for the averaged quantities $\langle Q_l^2 \rangle$. On large angular scales, corresponding to the lowest harmonics, only the Sachs-Wolfe effect operates. One has two terms corresponding to the scalar and tensor modes—we denote these contributions by square brackets. The scalar term is given in terms of the amplitude of the scalar density perturbation when it crosses the Hubble radius:

$$\left(\frac{\delta \rho}{\rho}\right)^{\rm HOR}_\lambda \equiv A_S \qquad (56)$$

by the integral

$$\Sigma_l^2[S] = \frac{8\pi^2}{m^2} \int_0^\infty \frac{dk}{k} j_l^2(2k/aH) A_S^2 T^2(k), \qquad (57)$$

where $j_l$ is a spherical Bessel function and the transfer function $T(k)$ is normalized to one on large scales.

The amplitude of a given Fourier mode of the dimensionless strain on scale $\lambda$ when it crosses inside the Hubble radius is given by

$$\left|k^{3/2} h_{\mathbf{k}}\right|^{\rm HOR}_\lambda \equiv A_G. \qquad (58)$$

The expression equivalent to $\Sigma_l^2[S]$ for the tensor modes contribution to temperature fluctuations is a rather complicated multiple integral which usually must be calculated numerically (Abbot and Wise, 1984; Starobinsky, 1985; Lucchin, *et al.* 1992). Under many circumstances (Lucchin, Matarrese and Mollerach suggest $0.5 < n < 1$ for power-law inflation) there is a helpful approximation which is that the ratio $\Sigma_l^2[S]/\Sigma_l^2[T]$ is independent of $l$ and given by

$$\frac{\Sigma_l^2[S]}{\Sigma_l^2[T]} = \frac{A_S^2}{A_G^2}. \qquad (59)$$

On the sky, one does not observe each contribution to the multipoles separately. As uncorrelated stochastic variables, the expectations add in quadrature to give

$$\Sigma_l^2 = \Sigma_l^2[S] + \Sigma_l^2[T]. \qquad (60)$$

There are two obstructions of principle. These are

- Even if one could measure the $\Sigma_l^2$ exactly, the last scattering surface being closed means one obtains only a discrete set of information—a finite number of the $\Sigma_l$ covering some effective range of scales. There will thus be an uncountably infinite set of possible spectra which predict exactly the same set of $\Sigma_l$.

- One cannot measure the $\Sigma_l^2$ exactly. What one can measure is a single realization, the $Q_l^2$. As a sum of $2l + 1$ gaussian random variables, $Q_l^2$ has a probability distribution which is a $\chi^2$ distribution with $2l + 1$ degrees of freedom, $\chi_{2l+1}^2$. The variance of this distribution is given by

$$\text{Var}[Q_l^2] = \frac{2}{2l+1} \langle Q_l^2 \rangle^2, \tag{61}$$

though one should remember that the distribution is not symmetric. Each $Q_l^2$ is a single realization from that distribution, when we really want to know the mean. From a single observer point, there is no way of obtaining that mean, and one can only draw statistical conclusions based on what can be measured. Thus, a larger set of spectra which give different sets of $\Sigma_l^2$ can still give statistically indistinguishable sets of $Q_l^2$. The variance falls with increasing $l$ but is significant right across the range of large scales.

## B. Intermediate scales ($\sim 200 h^{-1}$ Mpc $\longrightarrow 8 h^{-1}$ Mpc):

It is on intermediate scales that determination of the primeval spectrum is most promising. Here a range of promising observations are available, particularly towards the small end of the range of scales. In terms of technical difficulties in interpreting measurements, a trade-off has been made compared to large scales; on the plus side, the cosmic variance is a much less important player as far more independent samples are available, while on the minus side the spectrum has been severely affected by physical processes and thus has moved a step away from its primeval form.

1. Intermediate-scale microwave background anisotropies

In the absence of reionization, the relevant angular scales are from about 2° down to about 5 arcminutes. (Should reionization occur, a lot of the information on these scales could be erased or amended in difficult to calculate ways.) Several experiments are active in this range, including the South Pole and MAX experiments.

Unlike the large-scale anisotropy, one cannot write down a simple expression for the intermediate-scale anisotropies, even if it is assumed that one has already incorporated the effect of dark matter on the growth of perturbations via a transfer function. The reason is due to the complexity of physical processes operating. A case in point is the expected anisotropy (specified by the $\Sigma_l^2$, but now for larger $l$) in CDM models ($n = 1$), as calculated in detail by Bond and Efstathiou (Bond and Efstathiou, 1987).

On large scales, $l^2 \Sigma_l^2$ is approximately independent of $l$. Once we get into the intermediate regime, $l^2 \Sigma_l^2$ exhibits a much more complicated form, which is dominated by a strong peak at around $l = 200$. This is induced by Thomson scattering from moving electrons at the time of recombination. Bond and Efstathiou's calculation gives a peak height around 6 times as high as the extrapolated Sachs-Wolfe effect. Beyond the first peak is a smaller subsidiary peak at $l \sim 800$.

In their calculation, Bond and Efstathiou assumed both the primeval spectrum and the form of the dark matter. Of course, given the number of active and proposed dark matter search experiments, one should be optimistic that this information will be obtained in the not too distant future. However, even with this information, the complexity of the calculation makes it hard to conceive of a way of inverting it, should a good experimental knowledge of the $\Sigma_l^2$ ($l \in [30, 750]$) be obtained. Once again, it's much easier to compare a given theory with observation than to extract a theory from observation.

One of the interesting applications of these results might be a combination with the large-scale measurements. The peak on intermediate scales is due only to processes affecting the scalar modes, whereas we have pointed out that the large-scale Sachs-Wolfe effect is a combination of scalar and tensor modes. On large scales, one cannot immediately discover the relative normalizations of the two contributions. However, if the dark matter is sufficiently well understood, the height of the peak in the intermediate regime gives this information. Should it prove that the tensors do play a significant role, then this would be a very interesting result as it immediately excludes slow-roll potentials for the regime corresponding to the largest scales. Should the tensors prove negligible, then although the conclusions are less dramatic one has an easier inversion problem on large scales as one can concentrate solely on scalar modes.

2. Galaxy clustering in the optical and infrared

A. *Redshift surveys in the optical.*

Over the last decade, enormous leaps have been made in our understanding of the distribution of galaxies in the Universe from various redshift surveys. Most prominent is doubtless the ongoing Center for Astrophysics (CFA) survey (Ramella, Geller, and Huchra, 1992), which aims to form a complete catalogue of galaxy redshifts out to around $100h^{-1}$ Mpc. Other surveys of optical galaxies, often trading incompleteness for greater survey depth, are also in progress. On the horizon is the Sloan Digital Sky Survey which aims to find the redshifts of one million galaxies, occupying one quarter of the sky, with an overall depth of $500h^{-1}$ Mpc and completeness out to $100h^{-1}$ Mpc.

The redshifts of galaxies are relatively easy (though time consuming) to measure and interpret, and so provide one of the more observationally simple means of determining the distribution of matter in the Universe. The main technical problem is to correct the distribution for redshift distortions (which gives rise to the famous 'fingers of God' effect). However, the distribution of galaxies, specified by the galaxy power spectrum (or correlation function) is two steps away from telling us about the primeval mass spectrum.

- We have already discussed that the primeval spectrum on intermediate scales has been distorted by a combination of matter dynamics and amendments to the perturbation growth rate when the Universe is not completely matter dominated. If we know what the dark matter is, then this need not be a serious problem.

- Galaxies need not trace mass, and in modern cosmology it is almost always assumed they do not. This makes the process of getting from the galaxy power spectrum to the mass power spectrum extremely non-trivial. Models such as biased CDM rely on the notion of a scale-independent ratio between the two, but this too can only be an approximation to reality. In recent work, authors have emphasised the possible influence of environmental effects on galaxy formation (for instance, a nearby quasar might inhibit galaxy formation, and indeed it has been demonstrated that only very modest effects are required in order to profoundly affect the shapes of measured quantities such as the galaxy angular correlation function.

Despite this, attempts have been made to reconstruct the power spectrum from various surveys. In particular, this has been done for the CFA survey (Vogeley, *et al.* 1992), and for the Southern Sky Redshift Survey (Park, *et al.* 1992). These reconstructions remain very noisy, especially at both large scales (poor sampling) and small scales (shot noise and redshift distortions), and at present the best one could do would be to try and fit simple functional forms such as power-laws or parametrized power spectra to them. Even then, the constraints one would get on the slope of say a tilted CDM spectrum are very weak indeed. However, these reconstructions go along with the usual claim that standard CDM is excluded at high confidence due to inadequate large-scale clustering, without providing any particular constraints on the choice of methods of resolving this conflict.

Nevertheless, with larger sampling volumes such as those which the SDSS will possess, one should be able to get a good determination of the *galaxy* power spectrum across a reasonable range of scales, perhaps $10h^{-1}$ to $100h^{-1}$ Mpc.

### B. Redshift surveys in the infrared.

A rival to redshifts of optical galaxies is those of infra-red galaxies, based on galaxy positions catalogued by the Infra-Red Astronomical Satellite (IRAS) project in the mid-eighties. The aim here is to sparse-sample these galaxies and redshift the subset. This is being done by two groups, giving rise to the QDOT survey (Saunders, *et al.* 1991) and the 1.2 Jansky survey (Fisher, *et al.* 1992). Taking advantage of the pre-existing data-base of galaxy positions has allowed these surveys to achieve great depth with even sampling and reach some interesting conclusions.

The main obstacle to comparison with optical surveys is due to the selection method. Infra-red galaxies are generally young, and appear to possess a distribution notably less clustered than their optically selected counterparts. They are thus usually attributed their own bias parameter which differs from the optical bias. The mechanics of proceeding to the power spectrum are basically the same as for optical galaxies.

The most interesting and relevant results here are obtained in combination with peculiar velocity information, as discussed below.

### C. Projected catalogues.

As well as redshift surveys, one also has surveys which plot the positions of galaxies on the celestial sphere. At present the most dramatic is the APM survey, encompassing several million galaxies. The measured quantity is the projected counterpart of the correlation function, the angular correlation function usually denoted $w(\theta)$ where $\theta$ is the angular separation. Though arguments remain as to the presence of systematics, one in principle has accurate determinations of the galaxy angular correlation function. The first aim is to reconstruct the full three dimensional correlation function from this (proceeding thence to the galaxy power spectrum). Unfortunately, present methods of carrying out this inversion [based on inverting Limber's equation which gives $w(\theta)$ from $\xi(r)$] have proven to be very unstable, and a satisfactory recovery of the full correlation function has not been achieved.

In its preliminary galaxy identification stage, the SDSS will provide a huge projected catalogue on which further work can be carried out.

3. Peculiar velocity flows

Potentially the most important measurements in large-scale structure are those of the peculiar velocity field. Because all matter participates gravitationally, peculiar velocities directly sample the mass spectrum, not the galaxy spectrum. Were one to know the peculiar velocity field, this information is therefore as close to the primeval spectrum as is microwave background information.

Perhaps the most exciting recent development in peculiar velocity observations is the development of the POTENT method by Bertschinger, Dekel and collaborators (1989). Using only the assumption that the velocity can be written as the divergence of a scalar (in gravitational instability theories in the linear regime this is naturally associated with the peculiar gravitational potential), they demonstrate that the radial velocity towards/away from our galaxy (which is all that can be measured by the methods available) can be used to reconstruct the scalar, which can then be used to obtain the full three dimensional velocity field. This has been shown to work very well in simulated data sets, where one mimics observations and then can compare the reconstruction from those measurements with the original data set. So far, the noisiness and sparseness of available real radial velocity data has meant that attempts to reconstruct the fields in the neighbourhood of our galaxy have not yet met with great success; however, once better and more extensive observational data are obtained one can expect this method to yield excellent results.

At present, POTENT appears at its most powerful in combination with a substantial redshift survey such as the IRAS/QDOT survey. As POTENT supplies information as to the density field and the redshift survey to the galaxy distribution, the two in combination can be used in an attempt to measure quantities such as the bias parameter and the density parameter $\Omega_0$ of the Universe. Reconstructions of the power spectrum have also been attempted. At present, the error bars (due to cosmic variance because of small sampling volume, due to the sparseness of the data in some regions of the sky, and due to iterative instabilities) are large enough that a broad range of spectra (including standard CDM) are compatible with the reconstructed present-day spectrum.

With larger data sets and technical developments in the theoretical analysis tools, POTENT (and indeed velocity data in general) appears to be a very powerful means of investigating the present-day power spectrum. To that, one need only add a knowledge of the dark matter.

**C. Small scales ($8h^{-1}$ Mpc $\longrightarrow$ $1h^{-1}$ Mpc):**

It is worth saying immediately that this promises to be the least useful range of scales. For many choices of dark matter, including the standard hot dark matter scenario, perturbations on these scales are almost completely erased by dark matter free-streaming to leave no information as to the primeval spectrum. Only if the dark matter is cold does it seem likely that any useful information can be obtained.

There are several types of measurement which can be made. Quite a bit is known about galaxy clustering on small scales, such as the two-point galaxy correlation function. However, the strong nonlinearity of the density distribution on these scales erases information about the original linear-regime structure, and the requirement of $N$-body simulations to make theoretical predictions makes this an unpromising avenue for reconstruction even should nature have chosen to leave significant spectral power on these scales. There exist very small-scale (arcsecond–arcminute) microwave background anisotropy measurements, though these are susceptible to a number of line of sight effects, and further the anisotropies are suppressed (exponentially) on short scales because the finite thickness (about $7h^{-1}$ Mpc) of the last scattering surface comes into play.

Up to now, the most useful constraints on small scales have come from the pairwise velocity dispersion (the dispersion of line-of-sight velocities between galaxies). These are sensitive to the normalization of the spectrum at small scales, though unfortunately susceptible to power feeding down from higher scales as well. There are certainly noteworthy constraints—for instance it is generally accepted that unbiassed standard CDM generates excessively large dispersions. However, the calculations required involve $N$-body simulations and because a wide range of wavelengths contribute, obtaining knowledge of any structure in the power spectrum on these scales is likely to prove impossible, even if the amplitude can be determined to reasonable accuracy.

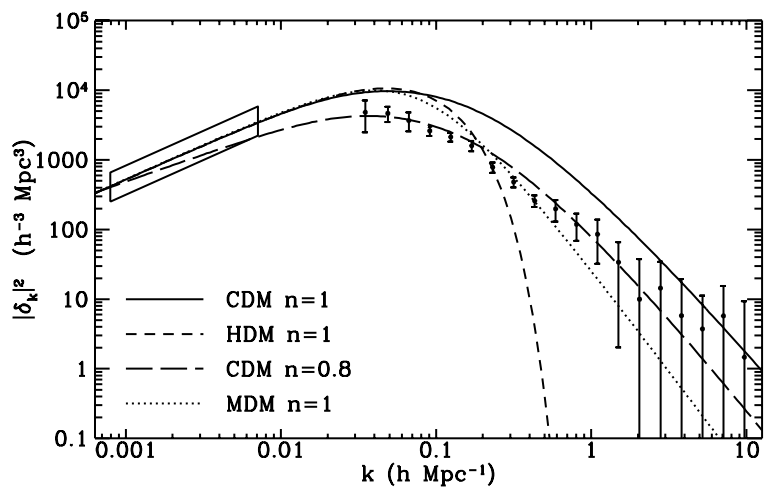

**Figure 6.** *Comparison of the measured power spectrum of density perturbations and the predictions of several models. The power spectrum of density perturbations from galaxies (the points on the right-hand side) is from Fisher, et al. 1992, and the power spectrum from the* COBE DMR *measurements (the box on the left-hand side) is from Smoot, et al. 1992. The models shown are cold dark matter; hot dark matter; tilted cold dark matter; and mixed dark matter.*

Already we are able to test various models for the power spectrum. For example a galaxy survey gives the power spectrum on 'small' scales, while CBR anisotropies probe the power spectrum on large scales. For preliminary models we can take a primordial power-law spectrum $|\delta_k|^2 \propto k^n$ ($n = 1$ is the Harrison-Zel'dovich spectrum) and some choice for dark matter, hot, cold, or mixed (fraction hot plus a fraction cold). Processing the primordial spectrum through the transfer function gives the curves of Figure 6. The simplest model is $n = 1$ CDM. Clearly the general shape is correct, but the normalization for the galaxy points does not match the normalization for the COBE points.

To better fit the observed spectrum several variations on the theme of CDM has been proposed. In the mixed dark matter (MDM) variant, a small amount of hot dark matter is added: $\Omega_{\rm HDM} \sim 0.3$, $\Omega_{\rm CDM} \sim 0.65$, and $\Omega_B \sim 0.05$. The 'pinch' of hot dark matter, *e.g.* in the form of neutrinos of mass 7 eV or so, leads to the suppression of fluctuations on small scales because of the free streaming of neutrinos. Another variant involves a modification of the spectrum of perturbations. When the spectrum of perturbations is normalized to the COBE DMR result, which fixes the spectrum on a very-large scale, a modest amount of 'tilt', say $n \simeq 0.8$, can reduce fluctuations on small scales. Another variant involves introducing a cosmological constant; a possibility too unpalatable to consider further.

Data on large-scale structure is accumulating rapidly. In a few years we will have much better information about the power spectrum, and we will see if any of the simple models is correct. Let us turn now to possibilities for generating the perturbations.

# 3  Inflation

The basic idea of inflation is that there was an epoch early in the history of the Universe when potential, or vacuum, energy was the dominant component of the energy density of the Universe. During that epoch the scale factor grew exponentially. During this phase (known as the de Sitter phase), a small, smooth, and causally coherent patch of size less than the Hubble radius $H^{-1}$ can grow to such a size that it easily encompasses the comoving volume that becomes the entire observable Universe today.

In the original proposal, inflation occurred in the process of a strongly first-order phase transition. This model was soon demonstrated to be fatally flawed. Subsequent models for inflation involved phase transitions that were second-order, or perhaps weakly first-order; some even involved no phase transition at all. Recently the possibility of inflation during a strongly first-order phase transition has been revived. Before discussing the latest developments in inflation, I will briefly review some of the history of inflation.

## 3.1  The art of inflation

**Pre-history**

The years before the birth of the inflationary Universe contained a rich pre-history of work in cosmology investigating the cosmological consequences of a Universe dominated by vacuum energy. Vacuum energy is interesting in cosmology because it acts as a cosmological constant, and will drive the Universe in exponential expansion. Recall that the expansion of the Universe is determined by the Friedmann equation:

$$\left(\frac{\dot{a}}{a}\right)^2 + \frac{k}{a^2} = \frac{8\pi G_N}{3}\rho, \tag{62}$$

where $a(t)$ is the cosmic scale factor, $\rho$ is the energy density of the Universe, and the constant $k$ is $\pm 1$ or $0$ depending upon the spatial curvature. If the contribution of the vacuum energy density $\rho_V$ dominates, then $\rho$ is a constant (it does not decrease with $a$), and for $k = 0$ the solution to the Friedmann equation is

$$a(t) = a(0)\exp(Ht); \qquad H \equiv \frac{\dot{a}}{a} = \left(\frac{8\pi G_N}{3}\rho_V\right)^{1/2} = \text{const.} \tag{63}$$

Such a rapid expansion may solve several cosmological problems, including the flatness/age problem, the homogeneity/isotropy problem, the problem of the origin of density inhomogeneities, and the monopole problem.

The possibility of a Universe dominated by vacuum energy became much more relevant with the realization that the Universe may have undergone a series of phase transitions associated with spontaneous symmetry breaking. The work of Kirzhnits and Linde (1972) showed that symmetries that are spontaneously broken today should have been restored at temperatures above the energy scale of spontaneous symmetry breaking, and as the Universe cooled below some critical temperature, denoted as $T_C$, there should have been a phase transition in which the symmetry was broken. Thus, phase transitions associated with spontaneous symmetry breaking might offer a mechanism whereby the early Universe may be dominated by vacuum energy for some period of time (Kolb and Wolfram, 1980).

**The Classical Era of Old Inflation**

Although there was a rich pre-history, the classical era of inflation crystallized with the paper of Guth (1980). In this classical picture, the Universe underwent a strongly first-order phase transition associated with spontaneous symmetry breaking of some Grand Unified Theory (GUT). Whether the phase transition is first order or higher order depends upon the details of the 'Higgs' potential for the scalar field whose vacuum expectation value is responsible for symmetry breaking. This theory is now usually referred to as 'old' inflation.

In old inflation the crucial feature of the potential was the barrier separating the symmetric high-temperature minimum, say located at $\phi = 0$, from the low-temperature true vacuum located at $\phi \neq 0$. If the transition is strongly first order, the transition from the high-temperature to the low-temperature minimum occurs by the quantum-mechanical process of nucleation of bubbles of true vacuum. These bubbles of true vacuum expand at the velocity of light, converting false vacuum to true.

The bubble nucleation rate ('per volume' will always be understood when discussing the bubble nucleation rate) depends upon the shape of the potential, but in general, it is written as $\Gamma = A\exp(-B)$, where $A$ is a parameter with mass dimension 4, and $B$, the bounce action, is dimensionless. Let us simply assume that $A \sim \phi_0^4$, where $\phi_0$ is the mass scale of spontaneous symmetry breaking (SSB).

In the classical picture, the energy density of the Universe became dominated by the false-vacuum energy of the Higgs field and the Universe expanded exponentially. Sufficient inflation was never a real concern; the problem with the classical picture is in the termination of the false-vacuum phase; usually referred to as the graceful exit problem.

Inside the true vacuum bubble is just what one expects—vacuum. For successful inflation it is necessary to convert the vacuum energy to radiation. The way this is accomplished in a first-order phase transition is through the process of collision of vacuum bubbles. In bubble collisions the energy density tied up in the bubble walls may be converted to entropy. Thus, if a first-order phase transition is to have a graceful exit, there must be many bubble collisions. The decline of the classical era began with the realization that bubbles of true vacuum do not percolate and fill the Universe; *i.e.* there is no graceful exit. The basic reason is that the exponential expansion of the background space overwhelms the bubble growth. To see this, consider the expression for the *coordinate* (or *comoving*) radius of the bubble. Assume that the bubble is nucleated at time $t_0$ with zero radius, and expands outward at the speed of light. At some time $t > t_0$ after nucleation, the comoving bubble radius is

$$r(t, t_0) = \int_{t_0}^{t} dt'\, a^{-1}(t') = \frac{\exp(-Ht_0) - \exp(-Ht)}{Ha(0)}. \tag{64}$$

The *physical* size of the bubble of course is simply $R(t, t_0) = a(t)r(t, t_0)$. Notice that as $t \to \infty$, the comoving bubble size approaches a *finite* value:

$$r(\infty, t_0) = \frac{\exp(-Ht_0)}{Ha(0)}. \tag{65}$$

Bubbles nucleated at larger $t_0$ reach a smaller comoving size than bubbles nucleated earlier in the transition. If a bubble is nucleated at time $t_0$, at some later time $t$ the bubble has comoving volume $v(t, t_0)$ and physical volume $V(t, t_0)$ given by

$$\begin{aligned} v(t, t_0) &= \frac{4\pi}{3} r^3(t, t_0) \longrightarrow \frac{4\pi}{3} \frac{\exp(-3Ht_0)}{[Ha(0)]^3} \\ V(t, t_0) &= \frac{4\pi}{3} R^3(t, t_0) \longrightarrow \frac{4\pi}{3} \frac{\exp[3H(t - t_0)]}{H^3}, \end{aligned} \tag{66}$$

where here arrows indicate the asymptotic values as $t \to \infty$.

The probability that a point remains in the old (false-vacuum) phase at time $t$ is simply

$$p(t) = \exp\left[-\int_0^t dt_0\, \Gamma V(t, t_0)\right] \longrightarrow \exp\left[-\frac{4\pi}{3}\left(\frac{\Gamma}{H^4}\right)Ht\right]. \tag{67}$$

Thus, the probability that a point remains in the false-vacuum phase decreases exponentially in time, just as expected.

Although $p(t)$ decreases exponentially, the volume of space in the false vacuum is increasing exponentially. A measure of whether true vacuum regions will percolate the space is the fraction of physical space in false vacuum:

$$f(t) = p(t)a^3(t) \longrightarrow \exp\left[-\frac{4\pi}{3}\left(\frac{\Gamma}{H^4}\right)Ht\right]\exp[3Ht]. \tag{68}$$

Clearly whether this fraction increases or decreases in time depends upon the competition between the decreasing probability for a point to be in the false vacuum and the increasing volume of space in the false vacuum. A rough estimate of whether $f(t)$ will increase or decrease is the criteria that $\epsilon = \Gamma/H^4$ is much greater or much less than unity. If $\epsilon$ is much less than one the transition will never be completed, while if $\epsilon$ is much greater than one the transition will be completed, but there won't be a sufficient period of inflation. So if $\epsilon$ is small enough to guarantee sufficient inflation, it will be too small for percolation to result.

This graceful exit problem led to the decline of the classical era of inflation and the dawn of the inflationary dark ages.

**Slow-Rollover Renaissance of New Inflation**

Soon after the demise of the original model, inflation was revived by the realization that it was possible to have an inflationary scenario without recourse to a strongly first-order phase transition. Linde, and Albrecht (1982) and Steinhardt (1982) proposed that the Universe inflates in the process of the classical evolution of the vacuum. In the classical evolution of the field to its true minimum the field has 'kinetic' energy and 'potential' energy. If one has a region of the scalar Higgs potential that is 'flat,' then the velocity of the Higgs field in the evolution to the ground state will be slow, and the potential energy of the Higgs field might dominate the kinetic energy. This can be made more quantitative by writing the classical equation of motion for a spatially homogeneous scalar field $\phi$ (called the *inflaton*) in an expanding Universe under the influence of a potential $V(\phi)$:

$$\ddot{\phi} + 3\frac{\dot{a}}{a}\dot{\phi} + \frac{dV(\phi)}{d\phi} = 0. \tag{69}$$

If the potential is flat enough that the $\ddot{\phi}$ term can be neglected, the scalar field will undergo a period of 'slow roll.' The energy density contributed by the scalar field is $\rho_\phi = \dot{\phi}^2/2 + V(\phi)$, and in the slow-roll region $V(\phi) \gg \dot{\phi}^2$, so the expansion closely approximates the exponential solution. This theory is sometimes referred to as 'new' inflation.

The original proposal of slow-rollover inflation was also based upon an SU(5) GUT phase transition. The potential was 'flattened' by assuming that it took the Coleman-Weinberg form. However it was soon realized that even this potential was not flat enough. If the scalar potential is approximated by a simple potential of the form $V(\phi) = \lambda(\phi^2 - \phi_0^2)^2$, in order for density fluctuations produced in inflation to be small enough required $\lambda \lesssim \mathcal{O}(10^{-15})$. Clearly such small numbers did not arise naturally in simple unified models, and a successful slow-rollover inflation model must be somewhat more complicated. Unfortunately, it was soon discovered that there is no cosmological upper bound on the complexity of models.

It was soon realized that the requirement of a small coupling constant could not easily be accommodated in simple particle physics models. Of course the usual temptation is to modify the Higgs sector by adding more representations than required in the minimal model. In fact a successful model was constructed along these lines by Pi (1984) and by Shafi and Vilenkin (1984).

For a while it was thought that supersymmetric GUTS could hold the key, but they were soon abandoned for a variety of reasons. After supersymmetric models, some very interesting supergravity models emerged. Although many supergravity models raised new problems of their own, some supergravity models were quite successful, and (at least) gave a proof of existence that the inflationary scenario might be implemented in particle models.

All of these Baroque models suffered from a low re-heat temperature as a result of a weakly coupled inflaton. This made baryogenesis problematical, although not impossible. All post-renaissance inflation models involved second-order transitions, and because inflation occurred in a smooth patch of the Universe that originally contained a single correlation region, the observable Universe should contain less than one topological defect produced in the transition. This is good news for the monopole problem, but bad news for cosmic strings and texture.

### Rococo Inflation

The complexity of inflationary models was again increased as people started modifying the gravitational sector of the theory. In Rococo inflation the identity of the inflaton is up for grabs. There are models where the inflaton is associated with the radius of internal dimensions, with the extra degree of freedom in fourth-order gravity, with the scalar field of induced gravity, *etc*. Some of these models can be made to work; it might be said that none work naturally.

Perhaps somewhere along the line as more and more detail was added to make the models satisfy all of the constraints, the message, or at least the spirit, of inflation was lost.

### Impressionism

In response to the excesses of Baroque and Rococo inflation, there grew up around Andrei Linde a Russian school of 'Impressionist' inflation. In the impressionist style no serious attempt is made to connect the details of inflaton with any specific particle physics models. In this way the true essence and beauty of the inflationary Universe is realized without any of the cluttering details. The best example of this the the 'chaotic' inflation model. In this model the scalar potential is assumed to be simply $V(\phi) = \lambda \phi^4$. What a perfect example of impressionism! This potential embodies features common to all scalar potentials without any of the details. Of course it is not 'realistic' in the sense that no one would accept the existence of a scalar field whose sole purpose is to make inflation simple, but it can be taken to represent the impressions of every scalar field, while at the same time representing no scalar field.

As Linde has repeatedly emphasized, it is not even necessary to connect inflation with a phase transition. In the $\lambda \phi^4$ chaotic model the $\phi$ field is expected to start away from its minimum (at $\phi = 0$) due to 'chaotic' initial conditions. From there, inflation can be analyzed as in slow rollover models.

Despite the seductive beauty of the impressionist approach we must demand more realism. Eventually we want a description of the Universe that has the fine details of the Baroque or Rococo but with the simplicity and spirit of impressionism.

**The Postmodern Era**

One of the most interesting recent developments is postmodernism. The postmodern movement is characterized by an eclectic mixture of classical tradition with some aspect of the recent past. With this definition, it may be said that first-order inflationary cosmologies represent a postmodern trend. The classical tradition is a first-order transition, while the aspect of the recent past will be embodied by the slow rolling of a second scalar field.

The key to first-order inflation is the relaxation of the assumption that $\epsilon \equiv \Gamma/H^4$ is constant in time. There are two ways one might imagine a time dependence for $\epsilon$. The first way is for $H$ to change. Since $H = \sqrt{G\rho_V}$, either the effective gravitational constant $G$ must change or the vacuum energy $\rho_V$ must change. (We will see that in many cases the two possibilities are equivalent representations of the same physics.) The second way is for $\Gamma$ to change. Of course, in general, both $H$ and $\Gamma$ might change.

If $\epsilon$ starts small, much less than one, then there might be a sufficient amount of inflation. If $\epsilon$ grows and eventually becomes much greater than one, then the bubbles of true vacuum will percolate and collisions between the bubble walls might convert the false-vacuum energy into entropy. This is the hope of first-order inflation.

The best example of a first-order inflation model is extended inflation (La and Steinhardt, 1989). The difference between extended inflation and Guth's model is the theory of gravity: Jordan-Brans-Dicke (JBD) in extended inflation rather than GR in Guth's model.

In JBD the gravitational 'constant' is set by the value of a scalar field. During inflation this scalar field evolves and gravity becomes weaker; as a result the cosmic-scale factor grows as a large power of time rather than exponentially. This means that in extended inflation the physical volume of space remaining in the false vacuum grows only as power of time and not exponentially, and unlike Guth's original model, bubble nucleation can convert all of space to the true vacuum.

## 3.2 Scalar field dynamics

Regardless of the particular model of inflation, scalar field dynamics plays an important role in the cosmology, so let us study the scalar field dynamics in more detail. Consider a minimally coupled, spatially homogeneous scalar field $\phi$, with Lagrangian density

$$\mathcal{L} = \frac{1}{2}\partial^\mu \phi\, \partial_\mu \phi - V(\phi) = \frac{1}{2}\dot\phi^2 - V(\phi). \tag{70}$$

With the assumption that $\phi$ is spatially homogeneous, the stress-energy tensor takes the form of a perfect fluid, with energy density and pressure given by $\rho_\phi = \dot\phi^2/2 + V(\phi)$, and $p_\phi = \dot\phi^2/2 - V(\phi)$. The classical equation of motion for $\phi$ is given in Equation 69. All minimal slow-roll models are examples of sub-inflationary behaviour, which is defined by the condition $\dot H < 0$. Super-inflation, where $\dot H > 0$, cannot occur here, though it is possible in more complex scenarios. This allows us to eliminate the time-dependence in the Friedmann equation and derive the first-order, non-linear differential equations

$$(H')^2 - \frac{3}{2}\kappa^2 H^2 = -\frac{1}{2}\kappa^4 V(\phi) \tag{71}$$

$$\kappa^2 \dot\phi = -2H', \tag{72}$$

where $\kappa^2 = 8\pi G$.

We can define two parameters, which we will denote as slow-roll parameters, by

$$\epsilon \equiv \frac{3\dot\phi^2}{2}\left(V + \frac{\dot\phi^2}{2}\right)^{-1} = \frac{2}{\kappa^2}\left(\frac{H'}{H}\right)^2$$

$$\eta \equiv \frac{\ddot\phi}{H\dot\phi} = \frac{2}{\kappa^2}\frac{H''}{H}. \tag{73}$$

Slow-roll corresponds to $\{\epsilon, |\eta|\} \ll 1$. With these definitions, the end of inflation is given *exactly* by $\epsilon = 1$. A small value of $\eta$ guarantees $3H\dot\phi \simeq -V'(\phi)$, which is often called the slow-roll equation.

Density perturbations arise as the result of quantum-mechanical fluctuations of fields in de Sitter space. First, let's consider scalar density fluctuations. To a good approximation we may treat the inflaton field $\phi$ as a massless, minimally coupled field. (Of course the inflaton does have a mass, but inflation operates when the field is evolving through a 'flat' region of the potential.) Just as fluctuations in the density field may be expanded in a Fourier series the fluctuations in the inflaton field may be expanded in terms of its Fourier coefficients $\delta\phi_\mathbf{k}$: $\delta\phi(\mathbf{x}) \propto \int \delta\phi_\mathbf{k} \exp(-i\mathbf{k}\cdot\mathbf{x}) d^3k$. During inflation there is an event horizon as in de Sitter space, and quantum-mechanical fluctuations in the Fourier components of the inflaton field are given by

$$k^3 |\delta\phi_\mathbf{k}|^2 / 2\pi^2 = (H/2\pi)^2, \tag{74}$$

where $H/2\pi$ plays a role similar to the Hawking temperature of black holes. Thus, when a given mode of the inflaton field leaves the Hubble radius during inflation, it has impressed upon it quantum mechanical fluctuations. In analogy to the discussion of the density perturbations of the previous section, what is called the fluctuations in the inflaton field on scale $k$ is proportional to $k^{3/2}|\delta\phi_\mathbf{k}|$, which by Equation 74 is proportional to $H/2\pi$. Fluctuations in $\phi$ lead to perturbations in the energy density $\delta\rho_\phi = \delta\phi(\partial V/\partial\phi)$.

Now considering the fluctuations as a particular mode leaves the Hubble radius during inflation, we may construct the gauge invariant quantity $\zeta$ using the fact that during inflation $\rho_0 + p_0 = \dot\phi^2$:

$$\zeta = \delta\phi\left(\frac{\partial V}{\partial\phi}\right)\frac{1}{\dot\phi^2}. \tag{75}$$

Now using Equation 71 and Equation 72, the amplitude of the density perturbation when it crosses the Hubble radius *after* inflation is

$$\left(\frac{\delta\rho}{\rho}\right)^{\text{HOR}}_\lambda \equiv \frac{m}{\sqrt{2}}A_S(\phi) = \frac{m\kappa^2}{8\pi^{3/2}}\frac{H^2(\phi)}{|H'(\phi)|} \propto \frac{V^{3/2}(\phi)}{m_{Pl}^3 V'(\phi)}, \tag{76}$$

where $H(\phi)$ and $H'(\phi)$ are to be evaluated when the scale $\lambda$ crossed the Hubble radius *during* inflation. The constant $m$ equals 2/5 or 4 if the perturbation re-enters during the matter or radiation dominated eras respectively. The 4 for radiation is appropriate to the uniform Hubble constant gauge. One occasionally sees a value 4/9 instead which is appropriate to the synchronous gauge. The matter domination factor is the same in either case. Note also that it is exact for matter domination, but for radiation domination it is only strictly true for modes much larger than the Hubble radius, and there will be corrections in the extrapolation down to the size of the Hubble radius.

Now we wish to know the $\lambda$-dependence of $(\delta\rho/\rho)_\lambda$, while the right-hand side of the equation is a function of $\phi$ when $\lambda$ crossed the Hubble radius during inflation. We may find the value of the scalar field when the scale $\lambda$ goes outside the Hubble radius in terms of the number of *e*-foldings of growth in the scale factor between Hubble radius crossing and the end of inflation.

It is quite a simple matter to calculate the number of *e*-foldings of growth in the scale factor that occur as the scalar field rolls from a particular value $\phi$ to the end of inflation $\phi_e$:

$$N(\phi) \equiv \int_{t_e}^{t} H(t')dt' = -\frac{\kappa^2}{2}\int_{\phi}^{\phi_e}\frac{H(\phi')}{H'(\phi')}d\phi'. \qquad (77)$$

The slow-roll conditions guarantee a large number of *e*-foldings. The total amount of inflation is given by $N_{\text{TOT}} \equiv N(\phi_i)$, where $\phi_i$ is the initial value of $\phi$ at the start of inflation (when $\ddot{a}$ first becomes positive). In general, the number of *e*-folds between when a length scale $\lambda$ crossed the Hubble radius during inflation and the end of inflation is given by

$$N(\lambda) = 45 + \ln(\lambda/\text{Mpc}) + \frac{2}{3}\ln(M/10^{14}\,\text{GeV}) + \frac{1}{3}\ln(T_{\text{RH}}/10^{10}\,\text{GeV}), \qquad (78)$$

where $M$ is the mass scale associated with the potential and $T_{\text{RH}}$ is the 're-heat' temperature. Relating $N(\lambda)$ and $N(\phi)$ from Equation 77 results in an expression between $\phi$ and $\lambda$. Hopefully this dry formalism will become clear in the example discussed below.

In addition to the scalar density perturbations caused by de Sitter fluctuations in the inflaton field, there are gravitational mode perturbations, $g_{\mu\nu} \to g_{\mu\nu}^{\text{FRW}} + h_{\mu\nu}$, caused by de Sitter fluctuations in the metric tensor. Here, $g_{\mu\nu}^{\text{FRW}}$ is the Friedmann-Robertson-Walker metric and $h_{\mu\nu}$ are the metric perturbations. That de Sitter space fluctuations should lead to fluctuations in the metric tensor is not surprising, since after all, gravitons are the propagating modes associated with transverse, traceless metric perturbations, and they too behave as minimally coupled scalar fields. The dimensionless tensor metric perturbations can be expressed in terms of two graviton modes we will denote as $h$. Performing a Fourier decomposition of $h$, $h(\vec{\mathbf{x}}) \propto \int \delta h_k \exp(-i\vec{\mathbf{k}}\cdot\vec{\mathbf{x}})d^3k$, we can use the formalism for scalar field perturbations simply by the identification $\delta\phi_{\mathbf{k}} \to h_{\mathbf{k}}/\kappa\sqrt{2}$, with resulting quantum fluctuations [*cf.* Equation 74]

$$k^3|h_{\mathbf{k}}|^2/2\pi^2 = 2\kappa^2(H/2\pi)^2. \qquad (79)$$

While outside the Hubble radius, the amplitude of a given mode remains constant, so the amplitude of the dimensionless strain on scale $\lambda$ when it crosses the Hubble radius after inflation is given by

$$\left| k^{3/2} h_{\mathbf{k}} \right|_{\lambda}^{\text{HOR}} \equiv A_G(\phi) = \frac{\kappa}{4\pi^{3/2}} H(\phi) \sim \frac{V^{1/2}(\phi)}{m_{Pl}^2}, \tag{80}$$

where once again $H(\phi)$ is to be evaluated when the scale $\lambda$ crossed the Hubble radius *during* inflation.

As usual, it is convenient to illustrate the general features of inflation in the context of the simplest model, chaotic inflation, which is to inflationary cosmology what *drosophila* is to genetics. In chaotic inflation the inflaton potential is usually taken to have a simple polynomial form such as $V(\phi) = \lambda\phi^4$, or $V(\phi) = \mu^2\phi^2$. For a concrete example, let us consider the simplest chaotic inflation model, with potential $V(\phi) = \mu^2\phi^2$. This model can be adequately solved in the slow-roll approximation, yielding

$$\begin{aligned}
\phi(t) &= \phi_i - \frac{2}{\sqrt{3}} \frac{\mu}{\kappa} t \\
a(t) &= a_i \exp\left[ \frac{\kappa\mu}{\sqrt{3}} \left( \phi_i t - \frac{\mu}{\sqrt{3}\kappa} t^2 \right) \right] \\
H &= \frac{\kappa\mu}{\sqrt{3}} \left( \phi_i - \frac{2}{\sqrt{3}} \frac{\mu}{\kappa} t \right) = \frac{\kappa\mu}{\sqrt{3}} \phi,
\end{aligned} \tag{81}$$

with inflation ending at $\kappa\phi_e = \sqrt{2}$ as determined by $\epsilon = 1$, where $\epsilon$ was defined in Equation 73. The number of $e$-foldings between a scalar field value $\phi$, and the end of inflation is just

$$N(\phi) = -\frac{\kappa^2}{2} \int_\phi^{\sqrt{2}/\kappa} \frac{H(\phi')}{H'(\phi')} d\phi' = \frac{\kappa^2 \phi^2}{4} - \frac{1}{2}. \tag{82}$$

Equating Equation 82 and Equation 78 relates $\phi$ and $\lambda$ in this model for inflation:

$$\kappa^2 \phi^2 / 4 = [45.5 + \ln(\lambda/\text{Mpc})]. \tag{83}$$

Using Equation 76 and Equation 80, $A_S$ and $A_G$ are found to be

$$\begin{aligned}
A_S(\lambda) &= \left( \sqrt{2}\kappa\mu/\sqrt{12\pi^3} \right) [45.5 + \ln(\lambda/\text{Mpc})] \\
A_G(\lambda) &= \left( \kappa\mu/\sqrt{12\pi^3} \right) [45.5 + \ln(\lambda/\text{Mpc})]^{1/2}.
\end{aligned} \tag{84}$$

We can note three features that are common to a large number of (but not all) inflationary models. First, $A_S$ and $A_G$ have different functional dependences upon $\lambda$. Second, $A_G$ and $A_S$ increase with $\lambda$. Finally, $A_S > A_G$, for scales of interest, although not by an enormous factor.

The basic picture of the generation of scalar and tensor perturbations is illustrated in Figure 7. The main observational information from the cosmic microwave background arises through the Cosmic Background Explorer (COBE) satellite, and the Tenerife (TEN) and South Pole (SP) collaborations. Galaxy surveys (APM, CFA, and IRAS) may provide useful information up to $100h^{-1}$ Mpc, while the Sloan Digital Sky Survey (SDSS) should extend to the lowest scales measured by COBE. Peculiar velocity measurements using the POTENT (P) methods are important on intermediate scales. The angle $\theta$ measures angular scales on the CBR in degrees, and length scales $\lambda$ are in units of $h^{-1}$ Mpc. $d_H$ refers to the horizon size today and at recombination and $d_{\text{NL}} \approx 8h^{-1}$ Mpc is the scale of non-linearity. (See the text for details).

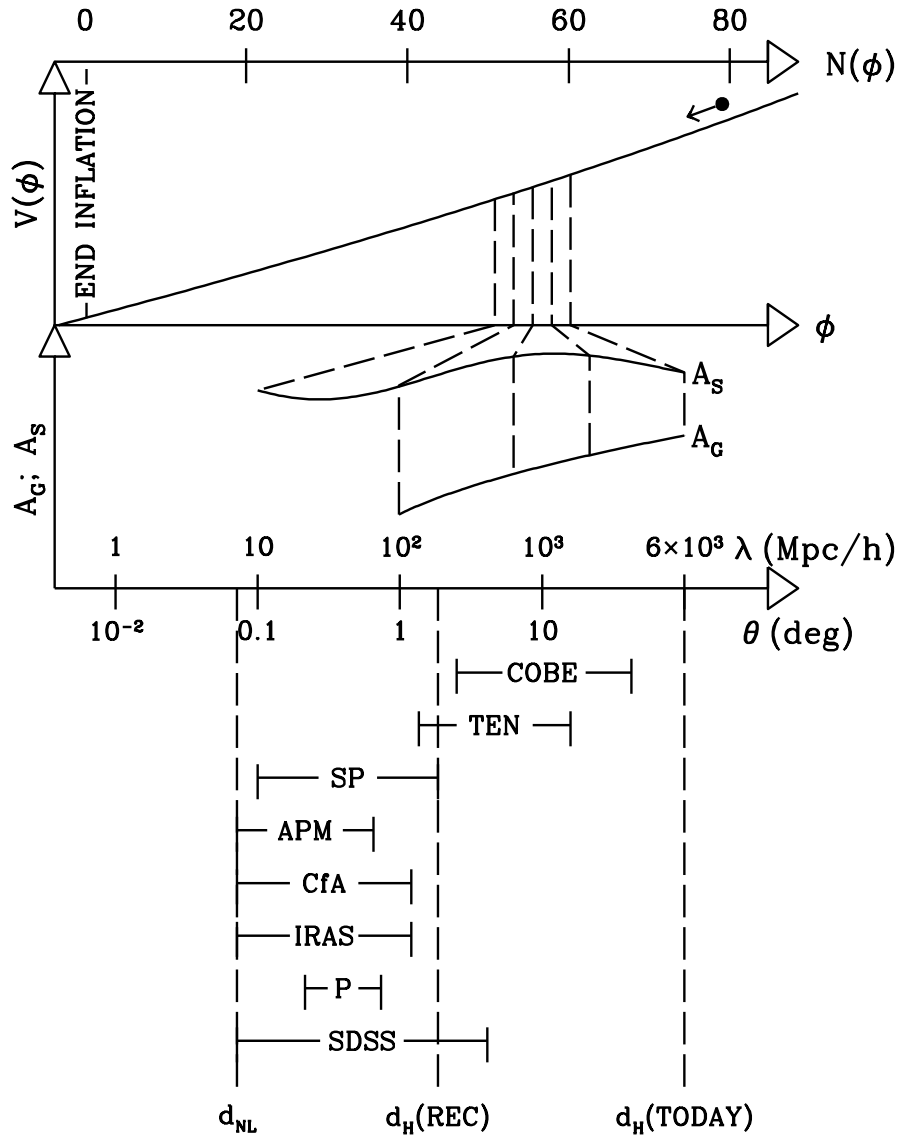

**Figure 7.** *A schematic figure illustrating the main concepts behind the generation of scalar and tensor perturbations in inflation .*

## 3.3 Reconstructing the Inflaton Potential

Figure 7 illustrates how a knowledge of the potential allows a prediction of the scalar and tensor spectra. Now let's consider the possibility of reverse engineering this approach and try to reconstruct the inflaton potential from knowledge of the scalar and tensor spectra.

Reconstruction of the inflaton potential in this manner was first considered by Hodges and Blumenthal (1990). Recently this question has been studied by Copeland, Kolb, Liddle, and Lidsey (CKLL) (1993a, 1993b, 1993c), and also by Turner (1993). CKLL improved upon the Hodges and Blumenthal (HB) results in two important ways. Firstly, they considered both scalar and tensor modes, whereas HB restricted their study to the scalars alone. This is a vital improvement, because, as HB realized, the scalar spectrum alone is insufficient to uniquely determine the inflaton potential—reconstruction is possible only up to an undetermined constant, and as the reconstruction equations are nonlinear, this leads to functionally different potentials giving rise to the same scalar spectrum. The tensors (even just the tensor amplitude at a single scale) provide just the extra information needed to lift this degeneracy. Secondly, the HB analysis made explicit use of the slow-roll approximation. It is well known that this approximation breaks down unless both the scalar spectrum is nearly flat and the tensor amplitude is negligible. CKLL considered the inflation dynamics in full generality. However, general expressions for the perturbation spectra were studied in a slow-roll expansion.

In CKLL (1993a), analytic expressions were derived for functionally reconstructing the potential in terms of $A_S$ and $A_G$. Although more complete expressions were given (see especially CKLL 1993c), here I will simply give the expressions to lowest order in $A_G/A_S$. The first result is a consistency equation relating the slope of the tensor spectrum to $A_G$ and $A_S$:

$$\frac{\lambda}{A_G(\lambda)} \frac{dA_G(\lambda)}{d\lambda} = \frac{A_G^2(\lambda)}{A_S^2(\lambda)}. \tag{85}$$

This highlights the asymmetry in the correspondence between the scalar and tensor spectra. If one were given the tensor spectrum, then a simple differentiation supplies the unique scalar spectrum. However, if a scalar spectrum is supplied, then this first-order differential equation must be solved to find the form of $A_G(\lambda)$. This leaves an undetermined constant in the tensor spectrum and, as the consistency equation is nonlinear, this implies that the scalar spectrum alone does not uniquely specify the functional form of the tensors. However, knowledge of the amplitude of the tensor spectrum at one scale is sufficient to determine this constant and lift the degeneracy.

It is the tensor spectrum one requires to proceed with reconstruction. Once the form of the tensor spectrum has been obtained, either directly from observation or by integrating the consistency equation, the potential, as parametrized by $\lambda$, may be derived:

$$V[\phi(\lambda)] = \frac{48\pi^3 A_G^2(\lambda)}{\kappa^4}, \tag{86}$$

where again this is true to lowest order in $A_G/A_S$. The reconstruction equations allow a functional reconstruction of the inflaton potential. For suitably simple spectra, this can be done analytically, and in CKLL 1993b we illustrated this for well-known cases of scalar spectra which are exactly scale-invariant, logarithmically corrected from scale-invariance, and exact power-laws. An alternative approach, useful for obtaining mass scales, is to concentrate on data around a given length scale $\lambda_0$, and perturbatively derive the potential around its corresponding scalar field value $\phi_0 \equiv \phi(\lambda_0)$. If we know $A_G(\lambda_0)$ and $A_S(\lambda_0)$ separately, then $V(\phi_0)$ follows immediately. The derivatives of the potential can also be obtained.

Let me illustrate the idea by example. Within a few years a combination of microwave background anisotropy measurements should give us some information about the scalar and tensor amplitudes at a particular length scale $\lambda_0$ (corresponding to an angular scale $\theta_0$). A hypothetical, but plausible, data set that this might provide would be $A_S(\lambda_0) = 1 \times 10^{-5}$; $A_G(\lambda_0) = 2 \times 10^{-6}$; $n_0 = 0.9$. This would lead to (see CKLL

1993a, 1993b)

$$V(\phi_0) = (2\times 10^{16}\text{GeV})^4; \quad \pm V'(\phi_0) = (3\times 10^{15}\text{GeV})^3; \quad V''_{\text{sr}}(\phi_0) = (5\times 10^{13}\text{GeV})^2. \quad (87)$$

In this way cosmology might be first to get a 'piece of the action' of GUT-scale physics.

# 4 Cosmological phase transitions

Perhaps the most important concept in modern particle theory is that of spontaneous symmetry breaking (SSB). The idea that there are underlying symmetries of nature that are not manifest in the structure of the vacuum appears to play a crucial role in the unification of the forces. In all unified gauge theories—including the standard electroweak model—the underlying gauge symmetry is larger than the unbroken SU(3)$_\text{C}\otimes$U(1)$_\text{EM}$. Of particular interest for cosmology is the theoretical expectation that at high temperatures, symmetries that are spontaneously broken today were restored, and that during the evolution of the Universe there were phase transitions associated with spontaneous breakdown of gauge (and perhaps global) symmetries. For example, we can be reasonably confident that there was such a phase transition at a temperature of order 300 GeV and a time of order $10^{-11}$ sec, associated with the breakdown of SU(2)$_\text{L}\otimes$U(1)$_\text{Y} \to$U(1)$_\text{EM}$. Moreover, the vacuum structure in many spontaneously broken gauge theories is very rich: topologically stable configurations of gauge and Higgs fields exist as domain walls, cosmic strings, and monopoles. In addition, classical configurations that are not topologically stable, so-called non-topological solitons, may exist and be stable for dynamical reasons. Interesting examples include soliton stars, Q-balls, non-topological cosmic strings, sphalerons, and so on.

Before discussing the cosmological implications, it is useful to review what is meant by the finite-temperature potential.

## 4.1 Finite-Temperature Potential

Let's start with a simple model of a real scalar field $\phi$ with Lagrangian

$$\mathcal{L} = \frac{1}{2}\partial_\mu\phi\,\partial^\mu\phi - V_0(\phi); \qquad V_0(\phi) = -\frac{1}{2}m^2\phi^2 + \frac{1}{4}\lambda\phi^4. \quad (88)$$

The Lagrangian is invariant under the discrete symmetry transformation $\phi \leftrightarrow -\phi$. The minima of the *classical* potential of Equation 88 are not at zero but at $\sigma_\pm = \pm\sqrt{m^2/\lambda}$. The origin, $\phi = 0$, is an unstable extremum of the potential because $V_0''(0) < 0$, where prime denotes $d/d\phi_c$. Since the quantum theory must be constructed about a stable extremum of the classical potential, the ground state of the system is either $\sigma_+$ or $\sigma_-$, and the reflection symmetry $\phi \leftrightarrow -\phi$ present in the Lagrangian is broken by the choice of a vacuum state, as $\phi = 0$ is the only possible vacuum invariant under $\phi \leftrightarrow -\phi$.

The potential of Equation 88 is the classical potential, and it is necessary to consider the effect of quantum corrections. Here I will follow the classic paper of Coleman and Weinberg (1973). For a general Lagrangian $\mathcal{L}(\phi)$ in the presence of a $c$-number source $J(x)$, the vacuum-to-vacuum amplitude is

$$\langle 0^+|0^-\rangle_J \equiv Z[J] = \int \mathcal{D}\phi \exp\left(i\int d^4x\,[\mathcal{L}(\phi) + J(x)\phi]\right), \quad (89)$$

where of course $Z[J]$ is the generating functional of the Green's functions. It is more useful to consider the generating functional of the *connected* Green's functions, $W[J]$, related to $Z[J]$ by $Z[J] = \exp(iW[J])$. $W[J]$ can be expanded in terms of powers of $J$, with the coefficients being the connected Green's functions. The classical field $\phi_c$ is defined as $\phi_c \equiv \delta W/\delta J$. Finally, the effective action is $\Gamma[\phi_c] = W[J] - \int d^4x J(x)\phi_c(x)$.

Now the effective action can be expanded in terms of $\Gamma^{(n)}$, the one-particle irreducible (1PI) Feynman diagrams with $n$ external lines:

$$\Gamma[\phi_c] = \sum_{n=1}^{\infty} \frac{1}{n!} \int \Gamma^{(n)}(x_1 \ldots x_n) \phi_c(x_1) \ldots \phi_c(x_n) d^4x_1 \ldots d^4x_n. \tag{90}$$

Rather than an expansion in powers of the classical field, one can expand the effective action in powers of derivatives of the classical field:

$$\Gamma[\phi_c] = \int d^4x \left[ -V(\phi_c) + \frac{1}{2} (\partial_\mu \phi_c)^2 Z(\phi_c) + \ldots \right]. \tag{91}$$

Now the constant term, $V(\phi_c)$, appearing in this expansion is known as the *effective potential*. By means of a Fourier transform of Equation 90, it is easy to show that the effective potential can also be expressed in terms of a sum of all 1PI Feynman graphs with zero momenta:

$$V(\phi_c) = -\sum_{n=1}^{\infty} \frac{1}{n!} \phi_c^n \, \Gamma^{(n)}(p_i = 0). \tag{92}$$

A simple example will illustrate the above dry formalism. In this example we follow Lee and Sciaccaluga (1975), and expand the effective potential about $\phi_c = \omega$, rather than about $\phi = 0$:

$$\begin{aligned}
\Gamma^{(n)}[\phi_c] &= \sum_{n=1}^{\infty} \frac{1}{n!} \int \Gamma^{(n)}(x_1 \ldots x_n) \, [\phi_c(x_1) - \omega] \ldots [\phi_c(x_n) - \omega] \, d^4x_1 \ldots d^4x_n, \\
V(\phi_c) &= -\sum_{n=1}^{\infty} \frac{1}{n!} [\phi_c - \omega]^n \, \Gamma^{(n)}(p_i = 0).
\end{aligned} \tag{93}$$

where the coefficients in the expansion are now the generators of the 1PI diagrams in the shifted theory where $\phi_c$ is replaced by $\phi_c - \omega$. Now $dV/d\omega|_{\phi_c = \omega} = \Gamma^{(1)}$, which is simply the tadpole diagram in the shifted theory (up to a factor of $i$). So evaluating the tadpole, integrating over $\omega$ then setting $\omega = \phi_c$ gives the effective potential.

The shifted theory of Equation 88 gives a potential with mass squared of $-m^2 + 3\lambda\omega^2$, and a $\phi^3$ term with coupling $3! \, i\lambda\omega$. Therefore the tadpole diagram is

$$\Gamma^{(1)} = \quad \raisebox{-0.3em}{\text{(tadpole diagram)}} \quad = -\frac{i}{2} \int \frac{d^4k}{(2\pi)^4} \frac{3! \, \lambda\omega}{k^2 - m^2 + 3\lambda\omega^2} \; .$$

The total potential to one-loop is the sum of the classical potential of Equation 88 and the one-loop correction:

$$V(\phi_c) = V_0(\phi_c) + \int^{\phi_c} \Gamma^{(1)} d\omega = V_0(\phi_c) + \frac{1}{2} \int \frac{d^4k}{(2\pi)^4} \ln \left( k^2 - m^2 + 3\lambda\phi_c^2 \right). \tag{94}$$

A generalization to arbitrary potentials would be to replace the last two terms in the logarithm by $V_0''(\phi_c)$, where $V_0$ is the classical potential.

A few comments are in order before proceeding.

1. The integral in Equation 94 is infinite. This shouldn't scare us. We introduce a cutoff $\Lambda$, and if the theory is renormalizable, all infinities can be absorbed via some renormalization prescription.

2. The physical meaning of the one-loop potential is clear if we integrate over $dk_0$ to find
$$V(\phi_c) = V_0(\phi_c) + \int \frac{d^3k}{(2\pi)^3} \sqrt{k^2 + V_0''(\phi_c)}. \tag{95}$$
Since $\sqrt{k^2 + V_0''(\phi_c)}$ corresponds to the total energy of a fluctuation of momentum $k$, the one-loop correction is clearly the sum of zero-point energy fluctuations about the point $\phi = \phi_c$.

3. If the model is generalized to include couplings of $\phi$ to vectors and fermions, then the one-loop potential will include additional tadpoles where the particle in the loop is the vector or the fermion.

Now the integral in Equation 94 can be done by introducing a cutoff $\Lambda$. For the simple model that we are studying, the potential to one-loop is

$$V(\phi_c) = -\frac{1}{2}m^2\phi_c^2 + \frac{1}{4}\lambda\phi_c^4 + \frac{(-m^2 + 3\lambda\phi_c^2)^2}{64\pi^2}\ln(-m^2 + 3\lambda\phi_c^2) + a_1(\Lambda)\phi_c^2 + a_2(\Lambda)\phi_c^4, \tag{96}$$

where $a_i(\Lambda)$ are cutoff-dependent constants that will be determined by renormalization of the mass and coupling constants.

Of course it is the behaviour of the theory at finite temperature that is of interest to us. A simple, heuristic derivation of the effects of the thermal bath is to adopt the 'real-time' formalism in which the propagator, $D(k)$ includes the possibility of emission and absorption from the thermal bath:

$$D_T(k) = \frac{1}{k^2 - m^2 + i\epsilon} + \frac{2\pi}{\exp(E/T) - 1}\delta(k^2 - m^2). \tag{97}$$

Then this propagator is used in the evaluation of the propagator of the tadpole diagram. The additional temperature-dependent part of the propagator leads to an additional, temperature-dependent part of the one-loop potential:

$$V_T(\phi_c) = \frac{T^4}{2\pi^2} \int_0^\infty dx\, x^2 \ln\left[1 - \exp\left(x^2 + V''(\phi_c)/T^2\right)^{1/2}\right]. \tag{98}$$

At high temperature, $T \gg |m|$, it is possible to expand the logarithm and perform the integration: $V_T(\phi_c) = -(\pi^2/90)T^4 + (1/24)V''(\phi_c)T^2 + \cdots$. The first term is simply the free energy of a massless spin-0 boson. The second term in the expansion is $\phi_c$-dependent, with a *positive* coefficient. For instance in the simple $\phi^4$ theory we have been following, $V''(\phi_c) = -m^2 + 3\lambda\phi_c^2$. Adding $V_T(\phi_c)$ to the classical potential gives a total potential with a coefficient of the term quadratic in $\phi_c$ of $-m^2/2 + \lambda T^2/8$. Clearly above some critical temperature $T_C = 2m/\lambda^{1/2}$ the coefficient of the quadratic term will be positive, and below $T_C$ the coefficient will be negative. This is a signal that for $T > T_C$ the symmetry will be restored, $\phi = 0$ will be a stable minimum of the potential. In the evolution from the high-temperature phase to the low-temperature phase there is a phase transition.

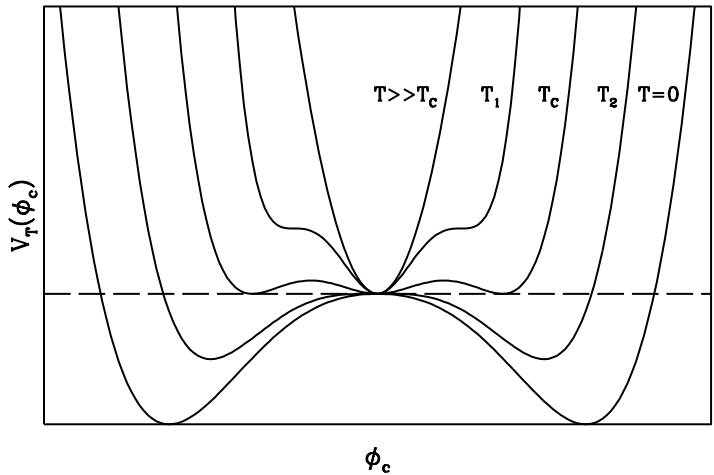

**Figure 8.** *The temperature dependence of $V_T(\phi_c)$ for a first-order phase transition. Only the $\phi_c$-dependent terms in $V_T(\phi_c)$ are shown.*

If $\phi$ also couples to fermions or gauge bosons, there will be additional terms in the temperature-dependent potential. The $\lambda \phi^4$ theory has a second-order transition. The additional terms in $V_T$ from gauge boson contributions can drive the transition to first order. In general, a symmetry-breaking phase transition can be first or second order. The temperature dependence of $V_T(\phi_c)$ for a first-order phase transition is shown in Figure 8. For $T \gg T_C$ the potential is quadratic, with only one minimum at $\phi_c = 0$. When $T = T_1$, a local mininum develops at $\phi_c \neq 0$. For $T = T_C$, the two minima become degenerate, and below $T_C$, the $\phi_c \neq 0$ minimum becomes the global minimum. If for $T \leq T_C$ the extremum at $\phi_c = 0$ remains a local minimum, there must be a barrier between the minima at $\phi_c = 0$ and $\phi_c \neq 0$. Therefore, the change in $\phi_c$ in going from one phase to the other must be discontinuous, indicating a first-order phase transition. Moreover, the transition cannot take place classically, but must proceed either through quantum or thermal tunnelling. Finally, when $T = T_2$ the barrier disappears and the transition may proceed classically. For a second-order transition there is no barrier at the critical temperature, and the transition occurs smoothly.

As a final illustration let's consider the electroweak phase transition. In the minimal electroweak model there is a complex SU(2) doublet $\Phi$, with potential $V(\Phi) = -m^2 \Phi^\dagger \Phi + \lambda_0 \left( \Phi^\dagger \Phi \right)^2$. Now the complex doublet field $\Phi$ can be expressed in terms of 4 real fields:

$$\Phi = \frac{1}{\sqrt{2}} \begin{pmatrix} \phi_1 & + & i\phi_2 \\ \phi & + & i\phi_3 \end{pmatrix}. \tag{99}$$

In the standard convention the vacuum expectation value of $\Phi$ is chosen to lie in the $\phi$ direction, $\langle \phi \rangle = \sigma$, and the real field $\phi$ has a potential like Equation 88. At tree level, the Higgs mass is $M_H^2 = 2\lambda_0\sigma^2$. The Higgs couples to gauge bosons, and lead to a mass for the $W$ and $Z$ of $M_W^2 = g^2\sigma^2/4$, $M_Z^2 = (g^2 + g'^2)\sigma^2$. The relationship between the mass of the $W$ and the Fermi constant gives $\sigma = 246$ GeV, $g = 0.66$ and $g' = 0.35$. The Higgs also couples to fermions, with Yukawa coupling $h_i = 0.57(M_i/100$ GeV$)$. As the Yukawa coupling is proportional to the fermion mass, the dominant effect is from the top quark, the most massive fermion. Therefore the 4 important coupling constants are $g = 0.66$, $g' = 0.35$, $h_T = 0.57(M_T/100$ GeV$)$, and $\lambda_0 = 0.08(M_H/100$ GeV$)^2$. Once the top quark and Higgs masses are determined, all the couplings in the minimal electroweak model will be known.

To the classical potential must be added the one-loop corrections from the Higgs, the $W$, the $Z$, and all the fermions (of course it is a good approximation to assume the fermion contributions are dominated by the top quark). At zero temperature, the one-loop effective potential is

$$V(\phi_c) = -\frac{1}{2}m^2\phi_c^2 + \frac{1}{4}\lambda\phi_c^4 + \frac{1}{64\pi^2}\big(-m^2 + 3\lambda\phi_c^2\big)^2 \ln\left(\frac{-m^2 + 3\lambda\phi_c^2}{\mu^2}\right)$$
$$+ \frac{3}{1024\pi^2}[2g^4 + (g^2 + g'^2)^2]\phi_c^4 \ln\left(\frac{\phi_c^2}{\mu^2}\right) - \frac{3}{64\pi^2}h_T^4\phi_c^4 \ln\left(\frac{\phi_c^2}{\mu^2}\right), \quad (100)$$

where $\mu$ is an arbitrary mass scale which can be related to the renormalized coupling constants. The gauge-boson contribution to the $\phi_c^4 \ln(\phi_c^2)$ term is $1.75 \times 10^{-4}$; the top-quark contribution is $-5.19 \times 10^{-4}(M_T/100$ GeV$)^4$; and the Higgs boson contribution is $9.73 \times 10^{-5}(M_H/100$ GeV$)^4$. A priori, all three contributions could be comparable. Let us consider the case where the Higgs mass is small ($M_H \lesssim 200$ GeV) and the Higgs contribution to the one-loop potential can be ignored. We can then write the potential of Equation 100 as

$$V(\phi_c) = -\frac{1}{2}m^2\phi_c^2 + \frac{1}{4}\lambda\phi_c^4 + B\phi_c^4 \ln\left(\frac{\phi_c^2}{\mu^2}\right)$$
$$= -\frac{1}{2}(2B + \lambda)\sigma^2\phi_c^2 + \frac{1}{4}\lambda\phi_c^4 + B\phi_c^4 \ln\left(\frac{\phi_c^2}{\sigma^2}\right). \quad (101)$$

Here we have used the fact that $V'(\sigma) = 0$ implies that $m^2 = (\lambda + 2B)\sigma^2$, and $B = 1.75$–$5.19 \times 10^{-4}(M_T/100\text{GeV})^4$. The Higgs mass is $M_H^2 = V''(\sigma) = 2(\lambda + 6B)\sigma^2$.

Now consider the potential at finite temperature. As in the previous example, the finite-temperature potential will have a temperature-dependent piece in addition to the zero-temperature part. The temperature-dependent part receives a contribution from all particles that couple to the scalar field, including the scalar field itself. The one-loop potential at finite temperature can be written as a sum of integrals similar to the one in Equation 98, of the form

$$F_\pm[X(\phi_c)] \equiv \pm \int_0^\infty dx\, x^2 \ln\left[1 \mp \exp[-(x^2 + X(\phi_c)/T^2)^{1/2}]\right] \quad (102)$$

($F_+$ applies to boson loops and $F_-$ to fermion loops). For the electroweak model, $V_T(\phi_c)$ is given by

$$V_T(\phi_c) = V(\phi_c) + \frac{T^4}{2\pi^2}\Big\{6F_+[g^2\phi_c^2/4] + 3F_+[(g^2 + g'^2)\phi_c^2/4]$$
$$+ F_+[M_H^2(\phi_c)] + 12F_-[h_T^2\phi_c^2/2]\Big\}, \quad (103)$$

where, as before, $V(\phi_c)$ is the one-loop potential at zero temperature, and for simplicity we have included only the $\phi_c$-dependent terms. For $T \gg \sigma$, the terms proportional to $T^4$ are just given by minus the pressure of a gas of the massless fermions and bosons that couple to $\phi$.

Anderson and Hall (1992) showed that a high temperature expansion of the one-loop potential closely approximates the full one-loop potential for $M_H \lesssim 150$ GeV and $M_T \lesssim 200$ GeV. (It is important to differentiate between the finite temperature Higgs mass, $M_H(T)$ and the zero-temperature Higgs mass, $M_H$.) They obtained for the potential

$$V(\phi) = D\left(T^2 - T_2^2\right)\phi^2 - ET\phi^3 + \frac{1}{4}\lambda_T\phi^4, \qquad (104)$$

where $D$ and $E$ are given by

$$D = \left[6(M_W/\sigma)^2 + 3(M_Z/\sigma)^2 + 6(M_T/\sigma)^2\right]/24; \quad E = \left[6(M_W/\sigma)^3 + 3(M_Z/\sigma)^3\right]/12\pi.$$

Here $T_2$ is given by

$$T_2 = \sqrt{(M_H^2 - 8B\sigma^2)/4D}, \qquad (105)$$

where the physical Higgs mass is given in terms of the one-loop corrected $\lambda$ as

$$M_H^2 = (2\lambda + 12B)\sigma^2, \text{ with } B = \left(6M_W^4 + 3M_Z^4 - 12M_T^4\right)/64\pi^2\sigma^4.$$

We use $M_W = 80.6$ GeV, $M_Z = 91.2$ GeV, and $\sigma = 246$ GeV. The temperature-corrected Higgs self-coupling is

$$\lambda_T = \lambda - \frac{1}{16\pi^2}\left[\sum_B g_B\left(\frac{M_B}{\sigma}\right)^4 \ln\left(M_B^2/c_B T^2\right) - \sum_F g_F\left(\frac{M_F}{\sigma}\right)^4 \ln\left(M_F^2/c_F T^2\right)\right] \qquad (106)$$

where the sum is performed over bosons and fermions (in our case only the top quark) with their respective degrees of freedom $g_{B(F)}$, and $\ln c_B = 5.41$ and $\ln c_F = 2.64$.

The gauge interactions result in an effective attractive $\phi^3$ term in the potential, so *to one-loop* the theory predicts a first-order electroweak phase transition. However the fact that the phase transition is so weakly first order implies that the finite-temperature loop expansion is not reliable, and the one-loop results cannot be trusted for the electroweak transition. There is presently a lot of work in finding an improved potential to describe the electroweak transition.

## 4.2　Generation of defects

SSB is an intergal part of modern particle physics, and provided that temperatures in the early Universe exceeded the energy scale of a broken symmetry, that symmetry should have been restored. How can we tell if the Universe underwent a series of SSB phase transitions? One possibility is that symmetry-breaking transitions were not 'perfect', and that false vacuum remnants were left behind, frozen in the form of topological defects: domain walls, strings, and monopoles.

As the first example of a topological defect associated with spontaneous symmetry breaking, consider the domain wall. The simple scalar model of the previous section can be used to illustrate domain walls. The $\mathcal{Z}_2$ reflection symmetry, *i.e.* invariance under $\phi \to -\phi$, of the Lagrangian of Equation 88 is spontaneously broken when $\phi$ takes on a non-zero vacuum expectation. So far we have assumed that all of space is in the same ground state, but this need not be the case! Imagine that space is divided into two regions. In one region of space $\langle \phi \rangle = +\sigma$, and in the other region of space $\langle \phi \rangle = -\sigma$. Since the scalar field must make the transition from $\phi = -\sigma$ to $\phi = +\sigma$ smoothly, there must be a region where $\phi = 0$, *i.e.* a region of false vacuum. This transition region between the two vacua is called a domain wall. Domain walls can arise whenever a $\mathcal{Z}_2$ (or any discrete) symmetry is broken.

The solution to the equation of motion, subject to the boundary conditions that describe a domain wall, is $\phi_W(z) = \sigma \tanh(z/\Delta)$, where the 'thickness' of the wall is characterized by $\Delta = (\lambda/2)^{-1/2}\sigma^{-1}$. It should be clear that the domain wall is topologically stable; the 'kink' at $z = 0$ can move around or wiggle, but it cannot disappear (except by meeting up with an antikink and annihilating). The stress tensor for the domain wall is obtained by substitution of the wall solution into the expression for the stress-energy tensor for a scalar field $T^\mu_\nu = (\lambda/2)\sigma^4 \cosh^{-4}(z/\Delta)\mathrm{diag}(1,1,1,0)$. Note that the $z$-component of the pressure vanishes, and that the $x$- and $y$-components of the pressure are equal to minus the energy density. The surface energy density associated with the wall, given by $\eta \equiv \int T^0_0 dz = (2\sqrt{2}/3)\lambda^{1/2}\sigma^3$, is identical to the integrated, transverse components of the stress, $\int T^i_i dz$. That is, the surface tension in the wall is precisely equal to the surface energy density. Because of this fact walls are inherently relativistic, and their gravitational effects are inherently non-Newtonian (and very interesting).

The existence of large-scale domain walls in the Universe today are ruled out simply based upon their contribution to the total mass density. A domain wall of size $H_0^{-1} \simeq 10^{28} h^{-1}$cm would have a mass of order $M_\mathrm{wall} \sim \eta H_0^{-2} \sim 4 \times 10^{65} \lambda^{1/2} (\sigma/100\mathrm{GeV})^3$ grams, or about a factor of $10^{10} \lambda^{1/2} (\sigma/100 \mathrm{\ GeV})^3$ times that of the total mass within the present Hubble volume. Walls would also lead to large fluctuations in temperature of the CBR unless $\sigma$ is very small: $\delta T/T \simeq G\eta H_0^{-1} \simeq 10^{10} \lambda^{1/2} (\sigma/100 \mathrm{\ GeV})^3$. Apparently, domain walls are cosmological bad news unless the energy scale and/or coupling constant associated with them are very small.

The existence of domain wall solutions for this simple model traces to the existence of the disconnected vacuum states: $\langle \phi \rangle = \pm \sigma$. The general mathematical criterion for the existence of topologically stable domain walls for the symmetry-breaking pattern $\mathcal{G} \to \mathcal{H}$ is that $\mathbf{\Pi_0}(\mathcal{M}) \neq \mathcal{I}$, where $\mathcal{M}$ is the manifold of equivalent vacuum states $\mathcal{M} \equiv \mathcal{G}/\mathcal{H}$, and $\mathbf{\Pi_0}$ is the homotopy group that counts disconnected components. In the above example, $\mathcal{G} = \mathcal{Z}_2$, $\mathcal{H} = \mathcal{I}$, $\mathcal{M} = \mathcal{Z}_2$, and $\mathbf{\Pi_0}(\mathcal{M}) = \mathcal{Z}_2 \neq \mathcal{I}$.

The next example of a topological defect is the cosmic string, a one-dimensional structure. As we shall see, cosmic strings are much more palatable to a cosmologist than domain walls. A simple model that illustrates the cosmic string is the Abelian Higgs model, a spontaneously broken U(1) gauge theory. The Lagrangian of the model contains a U(1) gauge field, $A_\mu$, and a *complex* Higgs field, $\Phi$, which carries U(1) charge $e$,

$$\mathcal{L} = D_\mu \Phi D^\mu \Phi^\dagger - \frac{1}{4} F_{\mu\nu} F^{\mu\nu} - \lambda (\Phi^\dagger \Phi - \sigma^2/2)^2, \tag{107}$$

where $F_{\mu\nu} = \partial_\mu A_\nu - \partial_\nu A_\mu$, and $D_\mu \Phi = \partial_\mu \Phi - ieA_\mu \Phi$. We immediately recognize that the theory is spontaneously broken as $V(\Phi)$ is minimized for $\langle |\Phi| \rangle^2 = \sigma^2/2$. The physical states after SSB are a scalar boson of mass $M_S^2 = 2\lambda\sigma^2$ and a massive vector boson of mass $M_V^2 = e^2\sigma^2$.

The complex field $\Phi$ can be written in terms of two real fields: $\Phi = (\phi + i\phi_1)/\sqrt{2}$. If the vacuum expectation value is chosen to lie in the real direction, then the potential becomes
$V(\phi) = (\lambda/4)(\phi^2 - \sigma^2)^2$, where $\langle |\Phi| \rangle = \langle \phi \rangle/\sqrt{2}$. However, energetics do not determine the phase of $\langle \Phi \rangle$ since the vacuum energy depends *only* upon $|\Phi|$; this fact follows from the U(1) gauge symmetry. Defining the phase of the vacuum expectation value by $\langle \Phi \rangle = (\sigma/\sqrt{2})\exp(i\theta)$, we see that $\theta = \theta(\vec{x})$ can be position dependent. However, $\Phi$ must be single valued; *i.e.* the total change in $\theta$, $\Delta\theta$, around any closed path must be an integer multiple of $2\pi$. Imagine a closed path with $\Delta\theta = 2\pi$. As the path is shrunk to a point (assuming no singularity is encountered), $\Delta\theta$ cannot change continuously from $\Delta\theta = 2\pi$ to $\Delta\theta = 0$. There must, therefore, be one point contained within the path where the phase $\theta$ is undefined, *i.e.* $\langle \Phi \rangle = 0$. The region of false vacuum within the path is part of a tube of false vacuum. Such tubes of false vacuum must either be closed or infinite in length, otherwise it would be possible to deform the path around the tube and contract it to a point without encountering the tube of false vacuum. In most instances, these tubes of false vacuum have a characteristic transverse dimension far smaller than their length, so they can be treated as one-dimensional objects and are called 'strings'.

The string solution to the equations of motion was first found by Nielsen and Olesen (1973). At large distances from an infinite string in the $z$-direction, their solution is

$$\Phi \longrightarrow (\sigma/\sqrt{2})\exp(iN\theta); \qquad A_\mu \longrightarrow -ie^{-1}\partial_\mu \left[\ln(\sqrt{2}\Phi/\sigma)\right], \qquad (108)$$

where $\theta$ is the polar angle in the $x$-$y$ plane, and $N$ is the winding number of the string. The stress-energy tensor associated with a long, thin, straight string is given by $T^\mu_\nu = \mu\delta(x)\delta(y)\text{diag}(1,0,0,1)$, where, $\mu$ is the mass per unit length of the string which depends upon the ratio $e^2/2\lambda$ but generally is of order $\pi\sigma^2$. Note that the pressure is negative—*i.e.* it is a string tension—and equal to $-\mu$. Like domain walls, strings are intrinsically relativistic.

Far from a circular string loop of radius $R$, the gravitational field is that of a point particle of mass $M_{\text{string}} = 2\pi R\mu$. For a loop of size about that of the present horizon, $M_{\text{string}} \simeq 10^{18}(\sigma/\text{GeV})^2$ grams. As with domain walls, there are non-Newtonian gravitational effects associated with strings. Recall that for a stress tensor of the form $T^\mu_\nu = \text{diag}(\rho, -p_1, -p_2, -p_3)$, the Newtonian limit of Poisson's equation is $\nabla^2\phi = 4\pi G(\rho + p_1 + p_2 + p_3)$. For an infinite string in the $z$ direction $p_3 = -\rho$ and $p_1 = p_2 = 0$, and Poisson's equation becomes $\nabla^2\phi = 0$, which suggests that space is flat outside of an infinite straight string. Indeed this is so. Vilenkin (1981) has solved Einstein's equations for the metric outside an infinite, straight cosmic string in the limit that $G\mu \ll 1$. In terms of the cylindrical coordinates $(r, \theta, z)$ the metric is $ds^2 = dt^2 - dz^2 - dr^2 - (1 - 4G\mu)^2 r^2 d\theta^2$. By a transformation of the polar angle, $\theta \to (1 - 4G\mu)\theta$, the metric becomes the flat-space Minkowski metric: as expected, space-time around a cosmic string is that of empty space. However, the range of the flat-space polar angle $\theta$ is only $0 \leq \theta \leq 2\pi(1 - 4G\mu)$ rather than $0 \leq \theta \leq 2\pi$. This is referred to as a conical singularity.

The conical nature of space around a string leads to several striking effects: double images of objects located behind the string, fluctuations in the microwave background, and the formation of wakes. To understand the formation of double images, consider the simplified situation of an infinite string normal to the plane containing the source and the observer. The conical space is flat space with a wedge of angular size $\Delta\theta$ removed and points along the cuts identified. Due to this, the observer will see two images of the source, with the angular separation, $\delta\alpha$, between the two images determined by

$$\sin(\delta\alpha/2) = \sin(\Delta\theta/2)\frac{l}{d+l}; \qquad \delta\alpha \simeq \Delta\theta\frac{l}{d+l} = 8\pi G\mu\frac{l}{d+l}. \qquad (109)$$

Here $l$ and $d$ are the distances from the string to the source and observer respectively, and the second equation is a small-angle approximation. The conical metric also leads to discontinuities in the temperature of the microwave background. Imagine as the source, a point on the last scattering surface for the microwave background radiation. An observer at rest with respect to the string will see two images of the same point on the last scattering surface, separated by an angle $\delta\alpha \simeq \Delta\theta$ (for $d \ll l$). Now if the string and observer are not at rest with respect to each other, but instead have a relative velocity $v$ which is perpendicular to the line of sight, the momentum vector of one image will have a small component (order $\Delta\theta$) parallel to the direction of $\vec{v}$, and the other, a small component antiparallel to the direction of $\vec{v}$. The net effect is a small Doppler shift of the radiation temperature $\delta T/T \simeq 8\pi G\mu v$ across the string. Based upon this effect and the observed isotropy of the CBR, we can conclude any strings that exist at present must be characterized by $G\mu \lesssim 10^{-5}$. A third interesting effect of cosmic strings are string wakes. Consider a long, straight string moving through the Universe with velocity $v$. As the string moves past particles in the Universe the particles will be deflected and will acquire a 'wake' velocity $v_W \simeq 4\pi G\mu v$, transverse to the direction of motion of the string. If the particles have a very small internal velocity dispersion, *e.g.* cold-dark matter particles, or baryons after decoupling, then matter on both sides of the passing string will move toward the plane defined by the motion of the string. In a Hubble time, a wedge-shaped sheet of matter, with overdensity of order unity, opening angle $\simeq 8\pi G\mu$, and width $vH^{-1}$, will form in the wake of the string. The mass of the material within the wake-produced sheet can be considerable, about $8\pi G\mu v^3$ of that in the horizon; likewise, the scale of the thin (thickness/width $\sim 8\pi G\mu$) sheets that are formed is comparable to that of the horizon scale. It has been suggested that the sheets that form in the wakes of long, straight cosmic strings play an important role in structure formation.

A final imprint of primordial cosmic strings is gravitational radiation from shrinking string loops. While an infinite straight string is stable, string loops are not. A curved string will move so as to minimize its length. The motion of a small, closed loop is particularly simple: a loop of radius $R$ oscillates relativistically, with a period $\tau \sim R$. As it oscillates it will radiate gravitational waves due to its time-varying quadrupole moment (dimensionally $Q \sim \mu R^3$). The power radiated in gravitational waves is given by $P_{GW} \simeq G(\dddot{Q})^2 \simeq \gamma_{GW} G\mu^2$ where $\gamma_{GW}$ is a numerical constant of order 100. In a characteristic time $\tau_{GW}$ the loop will radiate away its mass-energy, shrink to a point, and vanish. We expect $\tau_{GW} \sim \mu R/P_{GW} \sim (\gamma_{GW} G\mu)^{-1} R$. That is, a loop will undergo about $10^{-2}(G\mu)^{-1}$ oscillations before it disappears.

As cosmic strings stand, they are cosmologically safe and have several potentially interesting consequences: (1) they leave behind a background of relic gravitational waves; (2) relic string present today can lead to temperature fluctuations in the CBR; (3) relic string present today can act as gravitational lenses; and (4) string loops, or flattened structures formed in the wakes of strings, can possibly serve as seeds to initiate structure formation in the Universe.

In our discussion of cosmic strings, we have used the simplest example of a spontaneously broken gauge theory for which string solutions exist. In general, there will be string solutions associated with the symmetry breaking $\mathcal{G} \to \mathcal{H}$, if the manifold of degenerate vacuum states, $\mathcal{M} = \mathcal{G}/\mathcal{H}$, contains unshrinkable loops, i.e. if the mapping of $\mathcal{M}$ onto the circle is non-trivial. This is formally expressed by the statement that topologically stable string solutions exist if $\mathbf{\Pi_1}(\mathcal{M}) \neq \mathcal{I}$. In the above example $\mathcal{G} =$U(1), $\mathcal{H} = \mathcal{I}$, and $\mathcal{M} =$U(1). The group U(1) can be represented by the points on a circle, and so $\mathbf{\Pi_1}[\text{U}(1)]$ is the mapping of the circle onto itself. Such a mapping is characterized by the winding number of the mapping, i.e. $\theta \to N\theta$ ($N = 0, 1, \cdots$), so that $\Pi_1(\mathcal{M}) = \mathcal{Z}$, the set of integers.

Domain walls are two-dimensional topological defects, strings are one-dimensional defects. Point-like defects also arise in some theories which undergo SSB, and remarkably, they appear as magnetic monopoles. A simple model that illustrates the magnetic monopole solution is an SO(3) gauge theory, in which SO(3) is spontaneously broken to U(1) by a Higgs triplet $\Phi^a$, where $a$ is the group space index. The Lagrangian density for this theory is

$$\begin{aligned}
\mathcal{L} &= \frac{1}{2}D_\mu\Phi^a D^\mu\Phi^a - \frac{1}{4}F^a_{\mu\nu}F^{a\mu\nu} - \frac{1}{8}\lambda(\Phi^a\Phi^a - \sigma^2)^2, \\
F^a_{\mu\nu} &= \partial_\mu A^a_\nu - \partial_\nu A^a_\mu - e\varepsilon_{abc}A^b_\mu A^c_\nu, \\
D_\mu\Phi^a &= \partial_\mu\Phi^a - e\varepsilon_{abc}A^b_\mu\Phi^c.
\end{aligned} \quad (110)$$

Once again, we encounter a theory that undergoes SSB. In this model, two of the three gauge bosons in the theory acquire a mass through the Higgs mechanism. There is also a physical Higgs particle. The masses of the vector and Higgs bosons are $M_V^2 = e^2\sigma^2$, $M_S^2 = \lambda\sigma^2$.

The magnitude of $\langle\Phi^a\rangle$ is fixed by the minimization of the potential: $|\Phi| = \sigma$. However, the direction of $\langle\Phi^a\rangle$ in group space is not. This is just a manifestation of the SO(3) gauge symmetry. It should be clear that the lowest energy solution is the one where $\Phi^a(\vec{\mathbf{x}}) =$ const ($\vec{\mathbf{x}} =$ spatial coordinate) since this also minimizes the kinetic energy (spatial gradient term). Even if $\Phi^a(\vec{\mathbf{x}}) \neq$ const, the spatial dependence of the direction of $\Phi^a$ can often be gauged away, i.e. $D_\mu\Phi^a$ made equal to zero by an appropriate gauge configuration $A^a_\mu(\vec{\mathbf{x}})$, with finite energy. However, there are Higgs field configurations that cannot be deformed into a configuration of constant $\Phi^a$ by a finite-energy gauge transformation.

An example of a configuration that cannot be gauged away is the 'hedgehog' configuration, in which the direction of $\Phi^a$ in group space is proportional to $\hat{\mathbf{r}}$, where $\hat{\mathbf{r}}$ is the unit vector in ordinary space. This solution is spherically symmetric, and as $r$ tends to infinity, we find that $\Phi^a(r,t) \to \sigma\hat{\mathbf{r}}$, and $A^a_\mu(r,t) \to \varepsilon_{\mu ab}\hat{\mathbf{r}}_b/er$. Like the domain wall and the cosmic string solutions, continuity requires that the Higgs field vanish as $r \to 0$. The vanishing of the Higgs field at the origin accounts for the topological stability of the hedgehog: There is no way to smoothly deform the hedgehog into a configuration where $\langle |\Phi^a| \rangle = \sigma$ everywhere. The size of the monopole, *i.e.* the region over which $\langle |\Phi^a| \rangle \neq \sigma$, is of order $\sigma^{-1}$. The energy of the hedgehog configuration receives contributions from both the vacuum energy associated with $\langle |\Phi^a| \rangle \neq \sigma$ and spatial gradient energy associated with the variation of $\langle \Phi^a \rangle$.

Gauge and Higgs field configurations corresponding to a magnetic monopole exist if the vacuum manifold ($\mathcal{M} = \mathcal{G}/\mathcal{H}$) associated with the symmetry-breaking pattern $\mathcal{G} \to \mathcal{H}$ contains non-shrinkable surfaces, *i.e.* if the mapping of $\mathcal{M}$ onto the two-sphere is non-trivial. Mathematically, this is expressed by the statement that monopoles solutions arise in the theory if $\mathbf{\Pi_2}(\mathcal{M}) \neq \mathcal{I}$. If $\mathcal{G}$ is simply connected, then $\mathbf{\Pi_2}(\mathcal{G}/\mathcal{H}) = \mathbf{\Pi_1}(\mathcal{H})$. If $\mathcal{G}$ is not simply connected, then the generalization of the above expression is $\mathbf{\Pi_2}(\mathcal{G}/\mathcal{H}) = \mathbf{\Pi_1}(\mathcal{H})/\mathbf{\Pi_1}(\mathcal{G})$. In the example above, $\mathcal{G} = \mathrm{SO}(3)$, $\mathcal{H} = \mathrm{U}(1)$ (SO(3) is not simply connected—it is equivalent to the three-sphere with antipodal points identified), and $\mathbf{\Pi_2}[\mathrm{SO}(3)/\mathrm{U}(1)] = \mathbf{\Pi_1}[\mathrm{U}(1)]/\mathbf{\Pi_1}[\mathrm{SO}(3)] = \mathcal{Z}/\mathcal{Z}_2$, the integers mod 2.

We have discussed the three kinds of topological defects associated with spontaneously broken symmetries: the monopole; the string; and the domain wall. The existence and stability of these objects is dictated by topological considerations.

Many spontaneously broken gauge theories predict the existence of one or more of the above topological defects. These objects are inherently non-perturbative and probably cannot be produced in high energy collisions at terrestrial accelerators. It is very likely that the only place they can be produced is in phase transitions in the early Universe. Although monopoles, strings, and domain walls are topologically stable, they are not the minimum energy configurations. However, their production in cosmological phase transitions seems unavoidable. The 'unavoidable' cosmological production mechanism is known as the Kibble mechanism.

The Kibble mechanism hinges upon the fact that during a cosmological phase transition any correlation length is always limited by the particle horizon. The particle horizon is the maximum distance over which a massless particle could have propagated since the time of the bang. It was given in Equation 15. The correlation length associated with the phase transition sets the maximum distance over which the Higgs field can be correlated. The correlation length depends upon the details of the phase transition and is temperature-dependent. It is related to the temperature-dependent Higgs mass: $\xi \sim M_H^{-1}(T) \sim T^{-1}$. In any case, the fact that the horizon distance is finite in the standard cosmology implies that at the time of the phase transition ($t = t_C$, $T = T_C$), the Higgs field must be uncorrelated on scales greater than $d_H$, and thus the horizon distance sets an absolute maximum for the correlation length.

During a SSB phase transition, some Higgs field acquires a vacuum expectation value. Because of the existence of the particle horizon in the standard cosmology, when this occurs $\langle \phi \rangle$ cannot be correlated on scales larger than $d_H \sim H^{-1} \sim m_{Pl}/T^2$. Therefore, it should be clear that the non-trivial vacuum configurations must necessarily be produced, with an abundance of order one per horizon volume. While these topological creatures are not the minimum-energy configurations of the Higgs field, they arise as 'topological defects' because of the finite particle horizon. Since they are stable, they are 'frozen in' as permanent defects when they form.

So far we have concentrated on the analysis of topological defects that can arise in gauge theories. However defects can also arise in the spontaneous breaking of global symmetries. The analogies of the defects discussed above are global strings and global monopoles. The global field configurations look like their local counterparts for the scalar field, but of course there is no vector field. This means that formally the string and monopole solutions have infinite energy (recall for the local defects the energy in the gauge fields cancels the energy in the Higgs field far from the defect.) This is really not a problem, because there the divergence in the energy is only logarithmic, and there are many physical effects to cut it off (such as the inter-defect separation). There are just two main differences in the behaviour of gauge and global defects: (1) the energy of the global defects are slightly more spread out, (2) the global strings can radiate energy by the emission of Nambu-Goldstone bosons.

However there are new types of defects in global symmetry breaking that do not appear in the breaking of gauge symmetries. For example, in the spontaneous breaking of a global O($N$) model to O($N-1$), for $N = 1$ walls appear, for $N = 2$ global strings result, for $N = 3$ global monopoles are produced. These all have counterparts in local theories. However for $N > 3$ global defects also exist: for $N = 4$ the defect is called global texture, and for $N > 4$ they are called Kibble gradients. Texture corresponds to knots in the Higgs field that arise when the field winds around the three sphere. These knots are generally formed by misalignment of the field on scales larger than the horizon at the symmetry-breaking phase transition because of the Kibble mechanism. As the knots enter the horizon, they collapse at roughly the speed of light, giving rise to nearly spherical energy density perturbations. New knots are constantly coming into the horizon and collapsing, leading to a scale invariant spectrum of density perturbations. The magnitude of the perturbations is set by the scale of the symmetry breaking, and for scenarios of structure formation involving texture, the scale of symmetry breaking must be about $10^{16}$GeV.

A theory of texture or Kibble gradients being responsible for the seeds of large-scale structure has been formulated by Turok, Spergel and collaborators (Pen *et al.* 1993). Texture would provide a very promising alternative to conventional inflation scenarios for generating the primordial density fluctuations if indeed they are ubiquitous in particle physics models. In fact, texture arises in a variety of theories with non-Abelian global symmetries that are spontaneously broken. However, even an extremely small amount of explicit symmetry breaking will spoil the texture scenario. I would like to close these lectures by discussing how sensitive this theory is to Planck-scale effects. This idea was recently discussed by Holman *et al.* (1992) and Kamionkowski *et al.* (1992).

To illustrate these possibilities, consider a theory with a global O($N$) symmetry spontaneously broken to O($N-1$) by an $N$-vector. The theory is described by the scalar potential $V(\Phi) = \lambda \left(\Phi^a \Phi^a - \sigma^2\right)^2$. As mentioned above, texture arises for $N = 4$. There are many arguments suggesting that *all* global symmetries are violated at some level by gravity. For example, both wormholes and black holes can swallow global charge. 'Virtual' black holes or wormholes, which should, in principle, arise in a theory of quantum gravity, will lead to higher dimension operators which violate the global symmetry. There are two possible assumptions one might make about the fate of global symmetries in a Universe that includes gravity. The *strong* assumption is that, despite all indications from low-energy, semi-classical gravitational physics (black holes, wormholes, *etc.* ), it is possible to have exact global symmetries in the presence of gravity. This is the assumption made in the standard texture scenario. The *weak* assumption is that the global symmetry is not a feature of the full theory. There are two possible realizations of the weak assumption. Either the global symmetry is approximate, in which case one must include the effects of higher-dimensional, non-renormalizable, symmetry-breaking operators, or, consistent with indications from semi-classical quantum gravity, the global symmetry is never even an approximate symmetry unless protected by gauge symmetries.

If one makes the weak assumption, then one must include explicit symmetry breaking terms. If one assumes that gravity does not respect global symmetries at all, then renormalizable operators like $m_{Pl}^2 \lambda_{ab} \Phi^a \Phi^b$, which explicitly break the global symmetry, should be included. These terms are expected, for instance, by the action of wormholes swallowing global charge. If virtual wormholes of size smaller than the Planck length are included, then we expect $\lambda_{ab}$ to be of order unity. In this case it is wrong to consider an effective low-energy theory with a global symmetry. If one makes the assumption either that wormholes do not dominate the functional integral, or that the global charge is protected by gauge symmetries, then it may be possible to suppress the renormalizable operators. But even in this case higher dimension operators should be included. An example would be a dimension-5 operator, which would add to $V(\Phi)$ terms like $(\lambda_{abcde}/m_{Pl})\Phi^a \Phi^b \Phi^c \Phi^d \Phi^e$. Such terms explicitly break the global symmetry and lead to a mass for the pseudo-Nambu-Goldstone mode of $m^2 \propto \lambda \sigma^3/m_{Pl}$. Of course the mass is suppressed by $m_{Pl}$, but we will show below that it still has a drastic effect on the texture scenario.

The implications of the strong and weak assumptions for texture are as follows: With the strong assumption, the texture scenario is unaffected. If one allows unsuppressed wormhole contributions, global symmetries (and hence texture) are a non-starter. If all effects of gravitational physics in the low-energy theory are contained in non-renormalizable terms, a more careful analysis is required. This is the possibility we explore now. In this approach we are then required to include all higher dimension operators consistent with the gauge symmetries of the model and suppressed by appropriate powers of $m_{Pl}$.

We now consider the effects of the higher dimension operators. These terms will break the symmetry explicitly, generating a complicated potential for the Nambu-Goldstone modes. In general, the vacuum manifold will be reduced to a point, though the potential will likely have many local minima. To see how this works, consider the theory discussed above with $N = 3$. Here, the vacuum manifold is the two sphere and the model, in two spatial dimensions, will have texture. (In three spatial dimensions, the model admits both global monopoles and texture, although the texture in this case is not spherically symmetric. We express the field as

$$\Phi = \sigma(\sin\theta\cos\phi,\, \sin\theta\sin\phi,\, \cos\theta), \qquad (111)$$

where $\theta$ and $\phi$ are the angular variables on the two-sphere which represent the Nambu-Goldstone modes of the problem.

The effect of the dimension 5 operators is to introduce 21 terms to the potential for the field which depend explicitly on $\theta$ and $\phi$. (These are nothing more than the $Y_{1m}$, $Y_{3m}$, and $Y_{5m}$ spherical harmonics.) Note that in general, the mass of the Nambu-Goldstone boson in this potential is roughly $\sigma(\sigma/m_{Pl})^{1/2}$.

As long as the mass of the Nambu-Goldstone mode is small compared to the Hubble parameter, the field will evolve essentially as in the original texture scenario. However, once the Compton wavelength of the Nambu-Goldstone mode enters the horizon, the field will begin to oscillate about the minimum (or rather the closest local minimum) of its potential. The field will then align itself on scales larger than the horizon and texture on all scales quickly disappear. For texture to be important for structure formation, they must persist at least until matter-radiation decoupling when $H \simeq 10^{-28}$eV.

The contribution of a dimension $4+d$ operator to the Nambu-Goldstone boson mass is $m \sim \sigma(\sigma/m_{Pl})^{d/2}$. Given that the texture scenario requires $\sigma \sim 10^{16}$ GeV, the requirement that $m \lesssim 10^{-28}$eV implies that $d \gtrsim 35$; *i.e.* we must be able to suppress all operators up to dimension 40. It is rather difficult to see how this might occur; even the use of additional gauge quantum numbers could not prevent the occurrence of dimension 6 operators which break a non-Abelian symmetry (although they could protect an Abelian symmetry). We note that if we consider dimension-5 operators, then the mass becomes dynamically important immediately after the phase transition: texture therefore never exists.

In conclusion, any model which depends on the dynamics of Nambu-Goldstone modes will be extremely sensitive to physics at very high energies. Texture can by no means be considered a robust prediction of unified theories. This is most discouraging for the texture scenario. On the other hand, if texture is discovered, then this will have profound implications not only for theories of structure formation, but for Planck-scale physics. What better way to close lectures on the implications of cosmology for particle physics.

Finally there are other creatures that might be produced in cosmological phase transitions. Non-topological solitons, or Q-balls, (Frieman *et al.* 1988), and electroweak strings (Achucarro and Vachaspati, 1991).

The lesson for cosmological phase transitions is that even with unlimited energy, accelerators are the wrong tool to probe the non-perturbative sector of field theories. Early-Universe phase transitions continue to provide the best arena for the study of aspects of particle-physics theories related to coherent, soliton-like objects. The only plausible site for the production of objects such as monopoles, strings, walls, sphalerons, and the like is an early-Universe phase transition. All of these can have very significant implications for the evolution of the Universe. Sphalerons, as well as other solitons produced in the electroweak transition, have some promise of a cosmological payoff. Of course there is an enormous difference between finding a soliton-like solution to the field equations and finding solitons in the Universe. However, even if they are not are found, the techniques developed for their study will be useful additions to the theorist's toolbox.

# Acknowledgements

I am very grateful to several collaborators who greatly influenced the work recounted here: Ed Copeland, Marcelo Gleiser, Rich Holman, Andrew Liddle, Jim Lidsey, Mike Turner, Sharon Vadas, Yun Wang, and Erick Weinberg.

My work is supported in part by the Department of Energy and the National Aerounatics and Space Administration (Grant NAGW–2381) at at Fermilab.

In times of uncertainty about the future of our field of physics because of political and financial trouble, in times where the advances of our understanding of the Universe seemingly meet with hostility from an increasingly large segment of the public, in times when the prospects for young scientists seem grim, in times when the future of our field seems beyond our powers to influence, perhaps it is worthwhile to recall the words of Johannas Kepler in the last years of his life when his scientific career was caught in personal and political turmoil caused by the Thirty Years War:

> "When the storms rage around us, and the state is threatened by shipwreck, we can do nothing more noble than to lower the anchor of our peaceful studies into the ground of eternity."